\documentclass[aps,prx,twocolumn,nofootinbib,showpacs,showkeys,longbibliography,floatfix]{revtex4-2}

\usepackage{amsmath, amsxtra, amssymb, mathtools, isomath, nccmath, xspace, siunitx,graphicx,upgreek}
\usepackage{color,sidecap,natbib,dsfont,xcolor,pifont}\usepackage{hyperref}
\usepackage{upgreek}
\usepackage[utf8]{inputenc}
\usepackage[british]{babel}
\usepackage{hhline}
\usepackage{soul}
\usepackage{comment}
\usepackage{ragged2e}
\usepackage{braket}
\usepackage[normalem]{ulem}
\usepackage{placeins}
 
\hypersetup{ colorlinks=true, citecolor=blue, linkcolor=blue, urlcolor=blue }

\usepackage{newfloat}
\DeclareFloatingEnvironment[name={Extended Data Figure}]{extendedfigure}

\let\oldcite\cite
\renewcommand{\cite}[2][]{\unskip~\oldcite[#1]{#2}}

\begin{document}

\title{Continuous operation of a coherent 3{,}000-qubit system}

\author{
Neng-Chun~Chiu$^{1,*}$,
Elias~C.~Trapp$^{1,*}$,
Jinen~Guo$^{1,*}$,
Mohamed~H.~Abobeih$^{1,*}$,
Luke~M.~Stewart$^{1,*}$,
Simon~Hollerith$^{1,*,\dagger}$,
Pavel~L.~Stroganov$^{1}$, 
Marcin~Kalinowski$^{1}$, 
Alexandra~A.~Geim$^{1}$,
Simon~J.~Evered$^{1}$, 
Sophie~H.~Li$^{1}$,
Xingjian~Lyu$^{1}$,
Lisa~M.~Peters$^{1}$,
Dolev~Bluvstein$^{1}$,\\
Tout~T.~Wang$^{1}$, 
Markus~Greiner$^{1}$, 
Vladan~Vuleti\'{c}$^{2}$, and 
Mikhail~D.~Lukin$^{1, \dagger}$}

\affiliation{$^1$Department~of~Physics,~Harvard~University,~Cambridge,~MA~02138,~USA \quad \quad \\
$^2$Department~of~Physics~and~Research~Laboratory~of~Electronics,~Massachusetts~Institute~of~Technology,~Cambridge,~MA~02139,~USA\\
$*$ These authors contributed equally to this work.
}

\date{\today}

\begin{abstract}

Neutral atoms are a promising platform for quantum science, enabling advances in areas ranging from quantum simulations\cite{Browaeys2020, Gross2017, Greiner_2025} and computation\cite{Evered2023, atomcomputing2024, Endres2025, JThompson2025, saffman2025,Bluvstein2024,Manetsch2024} to metrology, atomic clocks\cite{Young2020, finkelstein2024universal, cao2024multi} and quantum networking\cite{covey2023quantum, hartung2024quantum, grinkemeyer2025error}.
While atom losses typically limit these systems to a pulsed mode, continuous operation\cite{Pause_2023,Norcia2024,Gyger2024, Bernien_2022, atomcomputing2025, Li2025} could substantially enhance cycle rates, remove bottlenecks in metrology\cite{ludlow2015optical}, and enable deep-circuit quantum evolution through quantum error correction\cite{BLuvstein2025,Baranes2025}. 
Here we demonstrate an experimental architecture for high-rate reloading and continuous operation of a large-scale atom-array system while realizing coherent storage and manipulation of quantum information.
Our approach utilizes a series of two optical lattice conveyor belts to transport atom reservoirs into the science region, where atoms are repeatedly extracted into optical tweezers without affecting the coherence of qubits stored nearby. 
Using a reloading rate of 300{,}000 atoms in tweezers per second, we create over 30{,}000 initialized qubits per second, which we leverage to assemble and maintain an array of over 3{,}000 atoms for more than 2 hours. 
Furthermore, we demonstrate persistent refilling of the array with atomic qubits in either a spin-polarized or a coherent superposition state while preserving the quantum state of stored qubits. Our results pave the way for the realization of large-scale continuously operated atomic clocks, sensors, and fault-tolerant quantum computers.

\end{abstract}
\maketitle

Neutral atom systems have recently emerged as a leading platform for quantum technologies, enabling advances in quantum simulations\cite{Browaeys2020,Gross2017}, quantum computing\cite{Evered2023, atomcomputing2024, Endres2025, JThompson2025, saffman2025, Bluvstein2024}, atomic clocks and metrology\cite{Young2020, finkelstein2024universal, cao2024multi}, and quantum networking\cite{covey2023quantum, hartung2024quantum, grinkemeyer2025error}.  However, an outstanding challenge associated with these systems involves atom loss, originating from errors in entangling operations\cite{Evered2023}, state-readout\cite{Bluvstein2024,BLuvstein2025}, and finite trap lifetime\cite{Zhang2024}. Atom losses necessitate pulsed operation which limits the performance of these quantum systems, including the circuit depth of quantum computation\cite{BLuvstein2025,Baranes2025}, accuracy of atomic clocks\cite{ludlow2015optical} and the rate of entanglement generation in quantum networking protocols\cite{pattison2024fast}. For instance in quantum computing, scaling up to large, practical algorithms requires encoding information in logical qubits, protected by repeated quantum error correction cycles\cite{Terhal2015,preskill_fault-tolerant_1997}. While these cycles can suppress error rates far below those of individual physical qubits\cite{Terhal2015,preskill_fault-tolerant_1997}, useful quantum circuits may require billions of operations, which eventually lead to loss of atomic qubits that need to be replaced\cite{Bluvstein2024,BLuvstein2025}. Similarly, atomic clock applications would benefit from coherent, continuous operation to improve duty cycles and reduce dead time, thereby enhancing stability and precision by mitigating Dick noise, one of the primary limitations of state-of-the-art optical clocks\cite{ludlow2015optical}. Addressing these challenges requires a reliable scheme for fast, continuous reloading of atomic qubits that not only outpaces the rate of errors due to decoherence and loss, but is also consistent with simultaneous coherent qubit storage and manipulation.

Recent experiments have enabled continuous atomic\cite{Biedermann2013} and optical clocks\cite{Schioppo2017, Katori_2024, Thompson_2025}, as well as the realization of continuous Bose-Einstein condensation\cite{Chen_2022}. While past efforts primarily focused on controlling atomic ensembles, most recently these techniques have been extended to explore continuous operation with individual atom control\cite{Bernien_2022, Pause_2023, Norcia2024, Gyger2024}. If expanded to high reloading rates within a coherence-preserving setting\cite{atomcomputing2025, Li2025}, these pioneering experiments highlight the exciting possibility of fully continuous operation of large-scale atomic systems.

\begin{figure*}[t!]
\centering
    \includegraphics[width=0.9\textwidth]{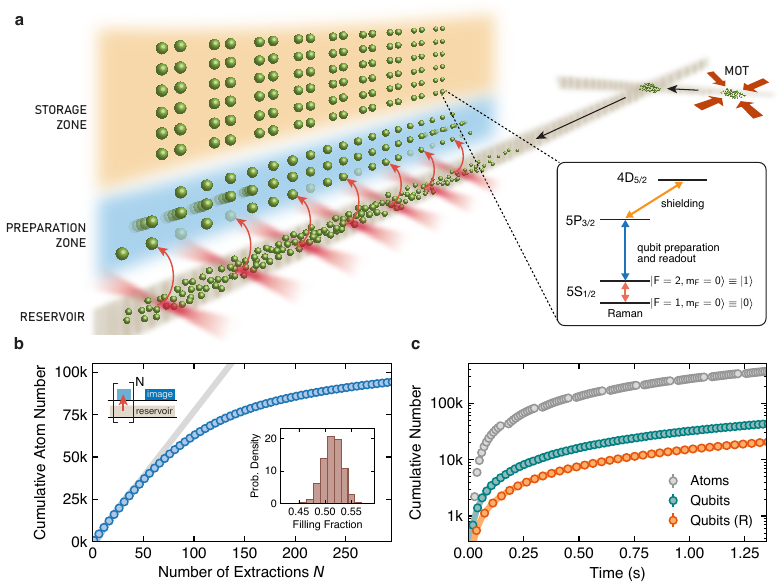}
    \caption{\justifying
    \textbf{Atom-array architecture for continuous operation.}
    \textbf{a,} 
    A cloud of laser-cooled atoms is transported over 0.5\,m from a separate MOT region into the science region via two optical lattice conveyor belts crossed at an angle. In the science region, the optical lattice serves as an atomic reservoir, from which a 2D array of optical tweezers repeatedly extracts atoms into the ``preparation zone". Here, atoms are laser-cooled, rearranged into a defect-free array and their qubit state initialized, then transferred into a large-scale storage tweezer array (``storage zone"). Our dual-lattice scheme avoids direct line-of-sight between the tweezer arrays and MOT location, and enables fully concurrent preparation and replenishment of the atomic reservoir. Inset: relevant atomic levels of $^{87}$Rb, where $F$ denotes the hyperfine level and $m_F$ the magnetic sublevel. During qubit preparation, storage qubits are protected from near-resonant photon scattering with the $5S_{1/2} \rightarrow 5P_{3/2}$ transition by light-shifting the excited state (``shielding"). Single-qubit gates are implemented via optical Raman transitions that drive clock states $\ket{0}$ and $\ket{1}$ (Methods).
    \textbf{b,} Cumulative number of atoms obtained by $N$-repeated tweezer extractions from a single lattice reservoir (see top-left schematic), where we observe a decline in tweezer filling fraction after $\sim\!70$ repeated extractions due to reservoir depletion (see also Extended Data Fig.\,\ref{fig:suppfig_twzloading}). For reference, the gray line indicates 50\% array filling. Inset: Histogram of tweezer filling fractions for the first 30 extractions from the reservoir. Notably, no laser cooling is applied during the tweezer loading process.
    \textbf{c,} Cumulative number of atoms/qubits obtained by tweezer extraction from repeatedly replaced lattice reservoirs. The gray markers indicate an atom flux of $\sim\!300{,}000$ atoms/s after light-assisted collisions, where the brief interruptions originate from the second transport stage of reservoir replacement during which no reservoir is present. Performing the qubit preparation sequence after each extraction, we achieve a continuous \textit{qubit} flux of 15{,}000 qubits/s with rearrangement (R; orange) and 30{,}000 qubits/s without rearrangement (green). Error bars represent the standard error of the mean across 10 repetitions.
    }
    \label{fig:fig1}
\end{figure*}

\begin{figure*}[t!]
\centering
\includegraphics[width=\textwidth]{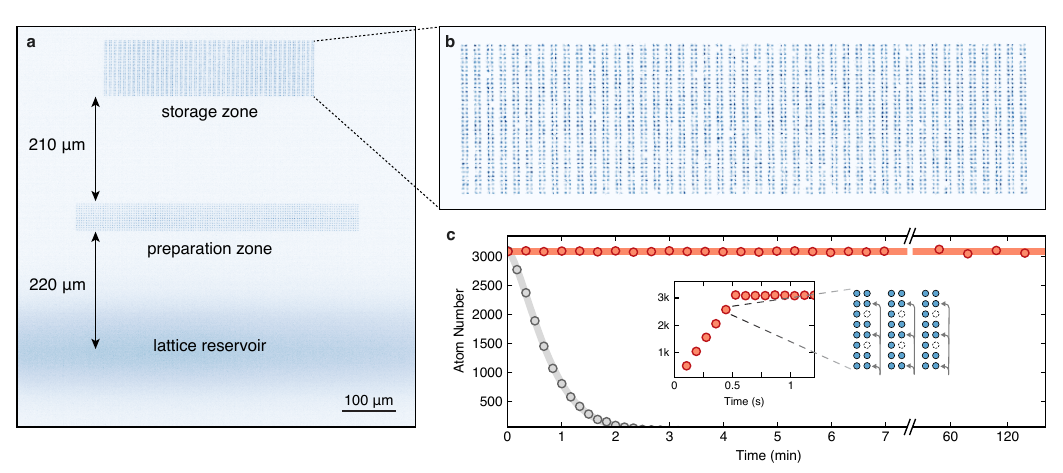}
    \caption{\justifying
    \textbf{Iterative assembly and continuous maintenance of a large-scale atomic array.}
    \textbf{a,} Atom fluorescence image outlining the zone architecture consisting of lattice reservoir, 1{,}440-site preparation zone, and 3{,}240-site storage zone. (Averaged) images of each zone are exposed separately and combined with different weights for visualization purposes.
    \textbf{b,} Single-shot fluorescence image of 3{,}217 atoms in the 3{,}240-site storage array (99.3\% filling).
    \textbf{c,} Iterative construction and continuous maintenance of a large-scale atomic array. Initial assembly occurs in $\approx\!0.5$\,s via six loading iterations (inset). Afterward, one of six segments (``subarrays") is ejected from the storage array and refilled with a fresh set of atoms every $\sim\!80\,$ms  (see also Extended Data Fig.\,\ref{fig:suppfig_rearrange} and Supplementary Video 1). Here, we show cyclic subarray replenishment and continuous maintenance of a 3{,}000+ atomic array for over 2 hours of operation, far beyond the tweezer-limited lifetime of $\sim\!60\,\mathrm{s}$ (gray). At the final datapoint, $t=2.3\,$h, over 50 million individually imaged and rearranged atoms have been cycled through the storage array. Error bars represent the standard error of the mean across 10 repetitions.
    }
\label{fig:fig2}
\end{figure*}

Here, we introduce a tweezer array architecture that enables such coherent continuous operation at large scale with reloading rates of up to 30{,}000 qubits per second, nearly two orders of magnitude above the current state of the art\cite{Norcia2024,Gyger2024}.
Our architecture is based on two serial optical lattice conveyor belts that transport a cloud of laser-cooled $^{87}$Rb atoms into the field-of-view of our microscope objective (Fig.\,\ref{fig:fig1}a). From this reservoir cloud, atoms are loaded into optical tweezers ``in the dark" (i.e., without laser cooling) and then repeatedly extracted into a ``preparation zone", where they are laser-cooled, imaged, rearranged, and initialized into their qubit states. Once initialized, atomic qubits are then transported and iteratively assembled into a large array in the ``storage zone", where dynamical decoupling is applied to maintain qubit coherence. Qubits in the storage zone are spatially protected against scattered cooling light by avoiding direct line-of-sight to the magneto-optical trap (MOT), and spectrally protected by light-shifting the cooling transition out of resonance (``shielding")\cite{Hu2024}. We demonstrate in-situ atom replenishment and maintenance of more than 3{,}000 storage array atoms for more than two hours, well beyond the trap lifetime of about 60\,s. Furthermore, we sustain the storage zone with either spin-polarized qubits ($Z$-basis) or qubits in the equal superposition state ($X$-basis) for, in principle, unlimited duration.

\section*{High-rate reloading from a lattice reservoir}

Our dual-lattice architecture is designed for uninterrupted high-rate qubit reloading that enables repetitive usage and periodic replacement of an atomic reservoir (Fig.\,\ref{fig:fig1}a). The experiment starts by loading around 4 million $^{87}$Rb atoms from a MOT into an optical lattice conveyor belt\cite{Wieman2000, Trisnadi2022, Klosterman2022, Matthies2024}. Then, the atom cloud is transported through a differential pumping tube to the separate science chamber, where it is transferred to a second lattice conveyor belt and delivered into the microscope field-of-view to serve as an atomic reservoir (Extended Data Figs.\,\ref{fig:suppfig_vac} and \ref{fig:suppfig_lattice}). Using this two-stage procedure, a fresh reservoir of $2.5$ million $120\,\upmu\mathrm{K}$ cold atoms arrives in the science region every $150\,\mathrm{ms}$. From the lattice reservoir, atoms are repeatedly loaded into a dynamic optical tweezer array of $120 \times 12$ sites, generated by a pair of crossed acousto-optic deflectors (AODs; Extended Data Fig.\,\ref{fig:suppfig_tweezers}). To load atoms, we switch on AOD tweezers overlapped with the lattice reservoir and immediately transport captured atoms into the preparation zone region placed $220\,\upmu\mathrm{m}$ above. This procedure takes less than $2.5\,\mathrm{ms}$ and, importantly, allows for multiple extraction cycles from a single reservoir. Following the extraction, atoms in AOD tweezers are transferred to a static tweezer array generated by a spatial light modulator (SLM).

Fig.\,\ref{fig:fig1}b demonstrates the results of repeated tweezer extraction from a single lattice reservoir. Here, we extract atoms for multiple cycles and only image and count single atoms after the final extraction cycle, and quote the cumulative number by summing the atom counts over all $N$ cycles. We find that, initially, array filling fractions of $>50\%$ are comparable to conventional tweezer loading from a MOT\cite{Grangier2002}, but gradually decline as the reservoir is depleted (see also Extended Data Fig.\,\ref{fig:suppfig_twzloading}). A key aspect of our dual-lattice design is the ability to extract atoms from one reservoir while preparing and delivering a fresh reservoir to the science chamber. By replacing reservoirs as they are depleted, this approach overcomes capacity limits of any single reservoir. In Fig.\,\ref{fig:fig1}c, we demonstrate this by repeatedly extracting atoms into the preparation zone as before, now replacing the reservoir every 60 tweezer loading cycles. As a result, we achieve a flux of $\sim\!300{,}000$ atoms in tweezers per second, corresponding to the maximum rate at which reservoir atoms can be extracted.

Notably, in contrast to the conventional approach to tweezer loading\cite{Grangier2002}, no laser cooling is applied during the extraction process. We attribute the ability to load optical tweezers ``in the dark" to a combination of stochastic overlap with atoms in the reservoir, and atomic collisions similar to the notion of a dimple trap\cite{Comparat2006} (Methods). While previous experiments\cite{Norcia2024, Gyger2024, Pause_2023} have relied on dissipative laser cooling or tweezer-lattice intensity ramps when loading fresh atoms from the reservoir, our scattering-free method helps preserve coherence of nearby storage qubits and avoiding lattice ramp-down enables repetitive usage of the reservoir.

\begin{figure*}[t!]
\centering
\includegraphics[width=1\textwidth]{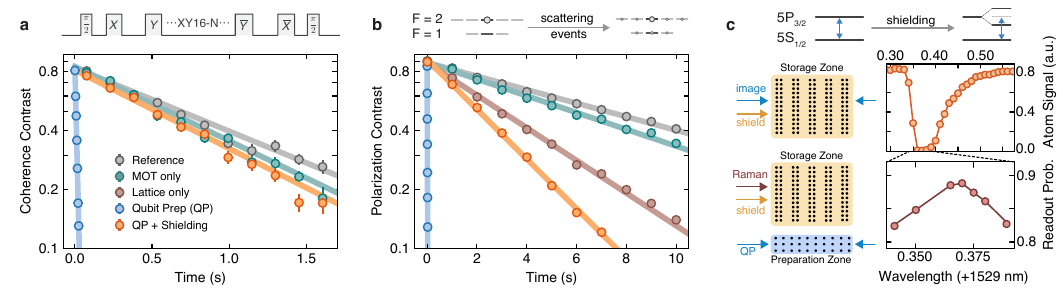}
    \caption{\justifying
    \textbf{Benchmarking concurrent qubit preparation.}
    \textbf{a,} Coherence contrast under various conditions when applying $N$-repetitions of an XY16 dynamical decoupling (DD) sequence with $\pi$-pulse spacing $2\tau \approx 1.6\,\mathrm{ms}$ to storage qubits, where the reference measurement yields $T_2 = 1.34(4)\,\mathrm{s}$ (gray). Operating the distant MOT in parallel to DD, we observe a minimal effect on coherence (green) compared to the reference, but a strong effect when additionally imaging in the preparation zone (blue). By applying qubit shielding, we restore coherence almost fully (orange, $T_2 = 1.09(3)\,\mathrm{s}$).
    \textbf{b,} A similar comparison probing depolarization of qubits initialized in $\ket{1}$ with the reference measurement $T_1 = 12.6(1)\,\mathrm{s}$ (gray), consistent with Raman scattering calculations owing to the tweezer light\cite{Moore2023}. While operating the MOT simultaneously has negligible effect on storage qubits (green), additionally imaging in the preparation zone results in rapid qubit depolarization (blue). Similar to before, this can be mitigated by shielding storage qubits from near-resonant light (orange, $T_1 = 3.43(3)\,\mathrm{s}$), mainly limited by off-resonant Raman scattering from the lattice light to which the shielding is ineffective (brown). A similar investigation for $\ket{0}$ state depolarization along with all measured $T_1$- and $T_2$-times is presented in Extended Data Fig.\,\ref{fig:supp_t1t2}. For subfigures (a-b), the difference of qubit populations measured in $\ket{0}$ and $\ket{1}$ provides the contrast (Methods).
    \textbf{c,} Shielding light spectroscopy on storage qubits. First, we image the storage array while applying low-power shielding light at variable wavelength to resolve the $4D_{5/2}$ resonance by suppression of imaging signal (top). In a fine-scan, we optimize for storage qubit coherence under DD while imaging in the preparation zone by maximizing the readout probability in $\ket{0}$ (bottom). Error bars represent the standard error of the mean across 10 repetitions.
    }
\label{fig:fig4}
\end{figure*}

To prepare atomic qubits, we perform an initialization procedure after every extraction from the reservoir (Extended Data Fig.\,\ref{fig:suppfig_stateprep}). Each step of this procedure relies on two counter-propagating laser beams local to the preparation zone and aligned coaxially with an externally applied static magnetic field (Methods)\cite{BLuvstein2025}. First, an explicit parity-projection pulse via finite-field polarization gradient cooling (PGC) on a red-detuned $F = 2 \rightarrow F' = 3$ transition prevents multiply-occupied optical tweezers. Here, $F$ denotes the hyperfine level of the atomic ground state and $F'$ the hyperfine level of the $5P_{3/2}$ excited state. We continue laser cooling via PGC during AOD-to-SLM handover, then apply a resonant pushout pulse to eliminate atoms in out-of-plane traps\cite{Manetsch2024}. This is followed by high-contrast, inherently background-free imaging (Methods). Afterward, we arrange atoms into a defect-free array while further laser cooling via electromagnetically induced transparency (EIT) with light blue-detuned from the $F = 2 \rightarrow F' = 2$ transition\cite{Kurtsiefer2024}. Finally, atoms are initialized to the qubit state $\ket{0}$ by optical pumping on the $F = 1 \rightarrow F' = 0$ transition, resulting in a state preparation and measurement (SPAM) fidelity of $\sim\!98\%$ within $20\,\upmu\mathrm{s}$. Under optimal conditions and without atom sorting, the qubit preparation sequence takes $20\,\mathrm{ms}$ (Methods).

Fig.\,\ref{fig:fig1}c shows the results of repeatedly extracting atoms and performing the qubit preparation sequence as described above, while the lattice reservoir is replaced in parallel every few tweezer extraction cycles. As a result, we achieve a \textit{qubit} flux of over $30{,}000$ qubits per second when choosing to not rearrange atoms. With atom sorting, the qubit preparation time approximately doubles and we obtain up to $15{,}000$ qubits per second, rearranged into defect-free batches of 600 qubits. In all cases, the qubit preparation time exceeds the time required for the second transport stage of reservoir replacement; as such, there is always a reservoir present for tweezer extraction and the qubit flux is uninterrupted.

\begin{figure*}[t!]
\centering
\includegraphics[width=1\textwidth]{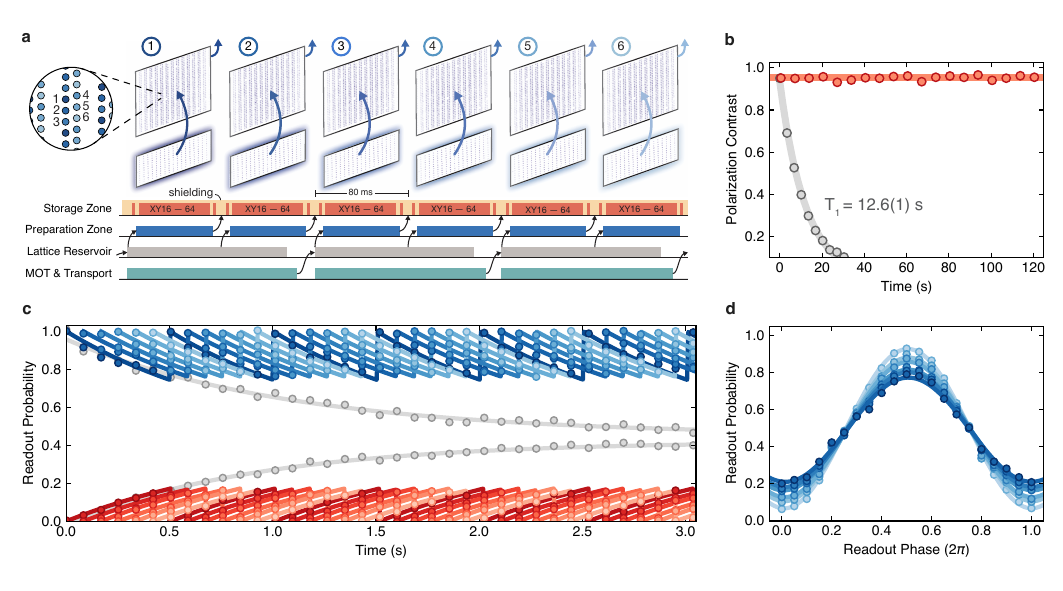}
    \caption{\justifying
    \textbf{Continuous operation while maintaining storage qubit coherence.}
    \textbf{a,} Time sequence visualizing our reloading protocol (see also Extended Data Fig.\,\ref{fig:contcohseq} and Supplementary Video 1). Following the initial storage array assembly, the longest-stored subarray is ejected and refilled with a preloaded set of qubits from the preparation zone every $\sim\!80\,$ms, while storage zone shielding is applied throughout. For subfigures (c-d), storage qubits are placed in the equal superposition state and undergo an XY16-64 decoupling sequence during each reloading cycle.
    \textbf{b,} Sequentially replenishing storage zone qubits, we maintain a high degree of storage array polarization (red) for, in principle, unbounded duration. For reference, we provide a $T_1$ measurement without qubit replenishment (gray).
    \textbf{c,} Similar to (b), now additionally applying an $\text{X}_{\pi/2}-(\text{XY16-64})-\text{X}_{-\pi/2}$ dynamical decoupling sequence during each subarray replenishment. We probe coherence of each subarray at various times during the replenishment cycle by reading out qubits in state $\ket{0}$ (blue) or $\ket{1}$ (red) as detailed in Methods. Individual subarrays (color shading) are unaffected by adjacent qubit reloading, and their dephasing is offset in time due to the cyclic subarray reloading protocol. The exponential sawtooth overlays are guides to the eye. For reference, we provide the $T_2$ measurement of a single subarray under the same cyclic decoupling sequence without qubit replenishment (gray).
    \textbf{d,} After multiple rounds of reloading under dynamical decoupling, we apply a final DD sequence and vary the phase of the last $\pi/2$-pulse to read out in different qubit bases. Complementary to (c), the observed coherence contrast varies for each subarray (color shading) due to the time-offset in subarray replenishment. Error bars represent the standard error of the mean across 10 repetitions.}
    \label{fig:fig3}
\end{figure*}

\section*{Assembly and maintenance of a large atom array}

After the preparation sequence, the rearranged array is transported to the storage zone which consists of 3{,}240 (90 $\times$ 36) SLM-generated optical tweezers with an average trap depth of $270\,\upmu\mathrm{K}$. The storage tweezer array features alternating regularly-spaced columns for lossless atom transport in between (Methods), and is positioned with sufficient distance to the preparation zone and the lattice reservoir to limit crosstalk between zones (Fig.\,\ref{fig:fig2}a).

We assemble the storage array in six iterations, each time transferring atoms into one of six segments (``subarrays") interspersed throughout the storage array (Extended Data Fig.\,\ref{fig:suppfig_rearrange}). Preparation and loading of each subarray, including atom transport to the storage zone, takes roughly $80\,\mathrm{ms}$ (mostly limited by rearrangement time and image data transfer). Assembly of the entire storage array is completed in $\approx\!500\,\mathrm{ms}$ with an averaged loading of 3{,}193 atoms (98.5\% filling fraction) over 300 trials. Fig.\,\ref{fig:fig2}b shows a single-shot fluorescence image of 3{,}217 atoms loaded into the array.

In Fig.\,\ref{fig:fig2}c, we demonstrate the ability to maintain over 3{,}000 atoms in the storage array for over two hours of continuous operation. After initial storage array assembly, we sequentially eject and refill the longest-stored subarray with a concurrently prepared set of fresh atoms from the preparation zone (see Supplementary Video 1). In parallel to atom preparation and subsequent replenishment, we replace the lattice reservoir every other tweezer extraction cycle without affecting the storage zone array. Using these techniques, we replenish atoms on much faster timescales than their tweezer-limited lifetime, and therefore enable operation that is, in principle, indefinite.

\section*{Coherence during continuous operation}

The ability to reload qubits while preserving coherence of existing qubits is essential for applications in deep-circuit quantum computation and high-bandwidth metrology\cite{Chen_2022,Savoie2018,BLuvstein2025}. To address this challenge, in Fig.\,\ref{fig:fig4}a we first investigate the impact of a simultaneously operating MOT on storage qubit coherence. We observe a coherence time of $T_2 = 1.15(3)\,$s when applying dynamical decoupling in the presence of the distant MOT, which shows minor modification compared to a reference measurement without the MOT ($T_2 = 1.34(4)\,$s). In Fig.\,\ref{fig:fig4}b, we find a similar result when probing storage qubit polarization via $T_1$ measurements with and without the MOT. Therefore, by preventing a direct line-of-sight between MOT and qubits, our angled dual-lattice transport scheme successfully disentangles the scattering-intense initial capture of an atomic gas from parallel quantum operations\cite{Ketterle2002, Katori_2024}.

In addition to the MOT, the coherence of existing qubits can be affected by scattered light and magnetic field changes during mid-circuit qubit preparation. To mitigate this, our beam architecture operates under constant finite magnetic field\cite{BLuvstein2025} and is localized to the preparation zone. However, we initially observe in Fig.\,\ref{fig:fig4}a that storage qubit coherence is strongly affected by beam crosstalk during the preparation zone imaging procedure. To suppress this effect, we protect qubits from near-resonant scattering by light-shifting the excited state\cite{Hu2024} as shown in Fig.\,\ref{fig:fig4}c, and find that the coherence time can be nearly completely restored ($T_2 = 1.09(3)\,$s). Additionally, we probe storage qubit depolarization under the same conditions in Fig.\,\ref{fig:fig4}b, resulting in a similar conclusion. Here, however, one observes an increased $T_1$-decay compared to a reference measurement despite shielding, which is largely dominated by off-resonant scattering from the lattice light. This increase does not measurably affect our $T_2$, but can be further mitigated by, for example, greater reservoir distance from the storage array, smaller lattice reservoir waist, or larger lattice detuning.

Building on these results, we now assess atom loss replenishment in simple quantum circuits by repeatedly replacing storage zone qubits while maintaining coherence. In Fig.\,\ref{fig:fig3}b, we first demonstrate high-rate reloading and continuous maintenance of a large array of spin-polarized storage qubits. Similar to Fig.\,\ref{fig:fig2}c, we now repeatedly prepare freshly initialized \textit{qubit} subarrays in the preparation zone, then eject and refill the oldest subarray in the storage zone as shown schematically in Fig.\,\ref{fig:fig3}a (see also Extended Data Fig.\,\ref{fig:contcohseq}). Sequentially replenishing qubits allows us to sustain a high degree of storage array polarization for, in principle, unbounded duration; here, we show maintenance of over 3{,}000 qubits for two minutes.

Finally, in Figs.\,\ref{fig:fig3}c,d we demonstrate the ability to reload and sustain a large array of atomic qubits in a coherent superposition state. While shielding and replenishing qubits as in Fig.\,\ref{fig:fig3}b, we additionally rotate storage qubits into state $\ket{+}$ and sustain coherence by applying a dynamical decoupling sequence during each subarray reloading cycle. Shortly before replenishing a qubit subarray with fresh qubits from the preparation zone, we map coherence into population by rotating all storage qubits into state $\ket{0}$, eject and replace the oldest qubits with newly spin-polarized ones, then rotate back into state $\ket{+}$ as a new reloading cycle starts. This enables us to keep qubits in a superposition state at about 90\% duty cycle, with the coherence of individual subarrays unaffected by concurrent reloading cycles.

\section*{Discussion and outlook}

Our experiments demonstrate a new atom-array architecture that enables continuous operation with reloading rates of up to 30{,}000 initialized qubits per second while preserving coherence across a rearranged large-scale qubit array. The results can be extended along several directions. First, the qubit preparation time can be significantly shortened through optimized readout and the use of FPGA-based and/or AI-optimized rearrangement protocols\cite{Wang2023, Lin2024}. Second, larger preparation zone arrays can be engineered by fully utilizing the system’s optical field-of-view. We estimate that these technical improvements would lead to a more than five-fold increase in qubit reloading rate, as this rate is directly proportional to qubit preparation time and preparation zone size. In addition, while the present experiments demonstrate continuous operation for over two hours, achieving much longer operation would benefit from active stabilization of the SLM-AOD tweezer overlap or automated beam alignment procedures. Finally, higher-power trapping lasers and high-efficiency diffractive optics, such as metasurfaces\cite{holman2024trapping}, can be immediately deployed to scale the storage and preparation zone size, supporting continuous operation of tens of thousands of atomic qubits.

Our results open up a range of new scientific opportunities based on atom-array platforms. In particular, our method is directly compatible with a zoned architecture for quantum computation involving Rydberg-mediated entangling gates, local optical Raman control, and dynamically reconfigurable qubit arrays\cite{Bluvstein2024,BLuvstein2025}. This architecture therefore presents a promising approach towards the implementation of deep, fault-tolerant quantum circuits using error correction. In a complementary experiment conducted in a separate apparatus, we demonstrate the core components of such a fault-tolerant quantum processor, including a method for mid-circuit loss-resolving qubit readout and re-use as well as deep-circuit protocols involving logical qubit teleportation, below-threshold repeated error correction, and universal fault-tolerant processing\cite{BLuvstein2025}. Atom losses play a major role in these experiments\cite{Baranes2025}, and the ultimate circuit depth is directly limited by atomic reservoir depletion. 

Taken together, our experiments open the door for realizing large-scale error-corrected quantum processors. For example, accounting for the current entangling gate fidelity ($\sim\!99.5\%$) and atom loss rate, at a 1\,ms duration per gate layer we estimate that 15{,}000 rearranged qubits per second should be sufficient to replenish lost atoms in a quantum processor with about 10{,}000 physical qubits. Furthermore, realistic improvements in entangling gate fidelities to $\sim\!99.9\%$ and reloading flux to 80{,}000 qubits per second could enable the operation of several hundred surface code logical qubits with a logical failure rate down to $10^{-8}$ \cite{Xu2024}. Moreover, the natural compatibility of this architecture with high-rate quantum low-density parity check (LDPC) codes will likely unlock further improvement in quantum processor performance\cite{Breuckmann2021,Bravyi2024}.

Beyond applications in quantum computation, a continuously operating atom-array system could overcome several limitations in quantum metrology\cite{ludlow2015optical}, enabling high-bandwidth and entanglement-enhanced precision quantum sensing\cite{cao2024multi, finkelstein2024universal}. Furthermore, a continuous stream of atomic qubits is essential to achieve fast generation of remote entanglement in quantum networking applications\cite{covey2023quantum, pattison2024fast, hartung2024quantum}. Finally, our high-rate reloading scheme and the transition from pulsed to continuous operation that it enables may be utilized to improve the performance of a broad class of cold-atom experiments, including quantum simulation, sensing and precision measurements.

\section*{Acknowledgments}
We thank Y. Bao, H. Bernien, S. Cantu, J. Dalibard, S. Ebadi, M. Endres, T. Esslinger, B. Grinkemeyer, N. Jepsen, S. Kolkowitz, J. Léonard, A. Lukin, T. Manovitz, J. Robinson, P. Sales Rodriguez, G. Semeghini, R. Tao, J. Ye, J. Zeiher and H. Zhou for discussions; and S. Geier, J. MacArthur and T.-K. Shen for help with the experiment. We acknowledge financial support from the US Department of Energy (DOE Quantum Systems Accelerator Center, contract number 7568717), IARPA and the Army Research Office, under the Entangled Logical Qubits program (Cooperative Agreement Number W911NF-23-2-0219), DARPA ONISQ program (grant number W911NF2010021) and MeasQuIT program (grant number HR0011-24-9-0359), the Center for Ultracold Atoms (an NSF Physics Frontier Center), the National Science Foundation (grant numbers PHY-2012023 and CCF-2313084) and QuEra Computing. M.H.A. acknowledges support by a Rubicon Grant from the Netherlands Organization for Scientific Research (NWO). S.H. acknowledges funding through the Harvard Quantum Initiative Postdoctoral Fellowship in Quantum Science and Engineering. S.J.E. acknowledges support from the National Defense Science and Engineering Graduate (NDSEG) fellowship. D.B. acknowledges support from the Fannie and John Hertz Foundation.\\

\noindent\textbf{Author contributions:} N.C.C., E.C.T., S.H., J.G., L.M.S., M.H.A., T.T.W. planned the experiments, performed measurements, analyzed the data, and contributed to building the experimental apparatus.
P.L.S., M.K. and X.L. contributed to the experimental control system. A.A.G., S.J.E., S.H.L. and D.B. contributed to development of methods and techniques for continuous operation. L.M.P. contributed to data interpretation. All work was supervised by M.G., V.V. and M.D.L. All authors discussed the results and contributed to the manuscript.\\

\noindent\textbf{Corresponding author:} Correspondence should be addressed to Mikhail D. Lukin (\href{mailto:lukin@physics.harvard.edu}{lukin@physics.harvard.edu}) and Simon Hollerith (\href{mailto:shollerith@fas.harvard.edu}{shollerith@fas.harvard.edu}).\\

\noindent\textbf{Competing interests:} M.G., V.V., and M.D.L. are co-founders and shareholders, V.V. is Chief Technology Officer, and M.D.L. is Chief Scientist of QuEra Computing. Some of the techniques and methods used in this work are included in provisional and pending patent applications filed by Harvard University (US patent application nos. 63/772,191 and 63/656,377).

\bibliography{Continuous_Reloading}

@Article{Terhal2015,
	author    = {Barbara M. Terhal},
	title     = {Quantum error correction for quantum memories},
	journal   = {Reviews of Modern Physics},
	year      = {2015},
	volume    = {87},
	number    = {2},
	pages     = {307--346},
	month     = {apr},
	doi       = {10.1103/revmodphys.87.307},
	publisher = {American Physical Society ({APS})},
}

@article{Moore2023,
  title = {{Photon scattering errors during stimulated Raman transitions in trapped-ion qubits}},
  author={{Moore, I.~D. \emph{et al.}}},
  journal = {Phys. Rev. A},
  volume = {107},
  pages = {032413},
  year = {2023},
  month = {Mar},
  publisher = {American Physical Society},
  doi = {10.1103/PhysRevA.107.032413},
  url = {https://link.aps.org/doi/10.1103/PhysRevA.107.032413}
}

@article{preskill_fault-tolerant_1997,
    title={Fault-tolerant quantum computation}, 
    author={Preskill, John},
    year={1997},
    eprint={quant-ph/9712048},
    journal={arXiv},
    primaryClass={quant-ph},
    url={https://arxiv.org/abs/quant-ph/9712048} 
}

@article{Bluvstein2024,
  author = {{Bluvstein, D. \emph{et al.}}},
  title = {Logical quantum processor based on reconfigurable atom arrays},
  journal = {Nature},
  year = {2024},
  day = {01},
  volume = {626},
  number = {7997},
  pages = {58-65},
  issn = {1476-4687},
  doi = {10.1038/s41586-023-06927-3},
  url = {https://doi.org/10.1038/s41586-023-06927-3}
}

@article{BLuvstein2025,
  author = {{Bluvstein, D. \emph{et al.}}},
  title={{A fault-tolerant neutral-atom architecture for universal quantum computation}},
  journal = {Nature},
  year = {2026},
  volume = {649},
  pages = {39-46},
  doi = {10.1038/s41586-025-09848-5},
  url = {https://doi.org/10.1038/s41586-025-09848-5}
}

@article{Zhang2024,
  title = {{High Optical Access Cryogenic System for Rydberg Atom Arrays with a 3000-Second Trap Lifetime}},
  author={{Zhang, Z. \emph{et al.}}},
  journal = {PRX Quantum},
  volume = {6},
  issue = {2},
  pages = {020337},
  numpages = {20},
  year = {2025},
  month = {May},
  publisher = {American Physical Society},
  doi = {10.1103/PRXQuantum.6.020337},
  url = {https://link.aps.org/doi/10.1103/PRXQuantum.6.020337}
}

@article{grinkemeyer2025error,
author = {{Grinkemeyer, B. \emph{et al.}}},
title = {Error-detected quantum operations with neutral atoms mediated by an optical cavity},
journal = {Science},
volume = {387},
number = {6740},
pages = {1301-1305},
year = {2025},
doi = {10.1126/science.adr7075},
URL = {https://www.science.org/doi/abs/10.1126/science.adr7075}}

@article{covey2023quantum,
author={Covey, Jacob P.
and Weinfurter, Harald
and Bernien, Hannes},
title={Quantum networks with neutral atom processing nodes},
journal={npj Quantum Information},
year={2023},
month={Sep},
day={16},
volume={9},
number={1},
pages={90},
issn={2056-6387},
doi={10.1038/s41534-023-00759-9},
url={https://doi.org/10.1038/s41534-023-00759-9}
}

@article{hartung2024quantum,
author = {Lukas Hartung  and Matthias Seubert  and Stephan Welte  and Emanuele Distante  and Gerhard Rempe },
title = {A quantum-network register assembled with optical tweezers in an optical cavity},
journal = {Science},
volume = {385},
number = {6705},
pages = {179-183},
year = {2024},
doi = {10.1126/science.ado6471},
URL = {https://www.science.org/doi/abs/10.1126/science.ado6471}}

@article{pattison2024fast,
  title={Fast quantum interconnects via constant-rate entanglement distillation}, 
  author={Christopher A. Pattison and Gefen Baranes and J. Pablo Bonilla Ataides and Mikhail D. Lukin and Hengyun Zhou},
  year={2024},
  eprint={2408.15936},
  journal={arXiv},
  primaryClass={quant-ph},
  url={https://arxiv.org/abs/2408.15936}, 
}

@Article{cao2024multi,
author = {{Cao, A. \emph{et al.}}},
title={{Multi-qubit gates and Schr{\"o}dinger cat states in an optical clock}},
journal={Nature},
year={2024},
month={Oct},
day={01},
volume={634},
number={8033},
pages={315-320},
issn={1476-4687},
doi={10.1038/s41586-024-07913-z},
url={https://doi.org/10.1038/s41586-024-07913-z}
}

@Article{finkelstein2024universal,
author = {{Finkelstein, R. \emph{et al.}}},
title={Universal quantum operations and ancilla-based read-out for tweezer clocks},
journal={Nature},
year={2024},
month={Oct},
day={01},
volume={634},
number={8033},
pages={321-327},
issn={1476-4687},
doi={10.1038/s41586-024-08005-8},
url={https://doi.org/10.1038/s41586-024-08005-8}
}

@Article{Young2020,
author = {{Young, A. \emph{et al.}}},
title={Half-minute-scale atomic coherence and high relative stability in a tweezer clock},
journal={Nature},
year={2020},
month={Dec},
day={01},
volume={588},
number={7838},
pages={408-413},
issn={1476-4687},
doi={10.1038/s41586-020-3009-y},
url={https://doi.org/10.1038/s41586-020-3009-y}
}

@article{ludlow2015optical,
  title = {Optical atomic clocks},
  author = {Ludlow, Andrew D. and Boyd, Martin M. and Ye, Jun and Peik, E. and Schmidt, P. O.},
  journal = {Rev. Mod. Phys.},
  volume = {87},
  issue = {2},
  pages = {637--701},
  numpages = {65},
  year = {2015},
  month = {Jun},
  publisher = {American Physical Society},
  doi = {10.1103/RevModPhys.87.637},
  url = {https://link.aps.org/doi/10.1103/RevModPhys.87.637}
}

@article{Norcia2024,
  title = {{Iterative Assembly of ${}^{171}$$\mathrm{Yb}$ Atom Arrays with Cavity-Enhanced Optical Lattices}},
  author = {{Norcia, M. \emph{et al.}}},
  journal = {PRX Quantum},
  volume = {5},
  issue = {3},
  pages = {030316},
  numpages = {13},
  year = {2024},
  publisher = {American Physical Society},
  doi = {10.1103/PRXQuantum.5.030316},
  url = {h}
}

@article{Ketterle2002,
author = {{Chikkatur, A. P. \emph{et al.}}},
title = {{A Continuous Source of Bose-Einstein Condensed Atoms}},
journal = {Science},
volume = {296},
number = {5576},
pages = {2193-2195},
year = {2002},
doi = {10.1126/science.296.5576.2193},
URL = {https://www.science.org/doi/abs/10.1126/science.296.5576.2193}}

@article{Pampel_2025,
  title = {{Quantifying Light-Assisted Collisions in Optical Tweezers across the Hyperfine Spectrum}},
  author = {Pampel, Steven K. and Marinelli, Matteo and Brown, Mark O. and D'Incao, Jos\'e P. and Regal, Cindy A.},
  journal = {Phys. Rev. Lett.},
  volume = {134},
  issue = {1},
  pages = {013202},
  numpages = {8},
  year = {2025},
  month = {Jan},
  publisher = {American Physical Society},
  doi = {10.1103/PhysRevLett.134.013202},
  url = {https://link.aps.org/doi/10.1103/PhysRevLett.134.013202}
}

@article{Pause_2023,
  title = {Reservoir-based deterministic loading of single-atom tweezer arrays},
  author = {Pause, Lars and Preuschoff, Tilman and Sch\"affner, Dominik and Schlosser, Malte and Birkl, Gerhard},
  journal = {Phys. Rev. Res.},
  volume = {5},
  issue = {3},
  pages = {L032009},
  numpages = {6},
  year = {2023},
  month = {Jul},
  publisher = {American Physical Society},
  doi = {10.1103/PhysRevResearch.5.L032009},
  url = {https://link.aps.org/doi/10.1103/PhysRevResearch.5.L032009}
}

@article{Gyger2024,
  title = {Continuous operation of large-scale atom arrays in optical lattices},
  author = {{Gyger, F. \emph{et al.}}},
  journal = {Phys. Rev. Res.},
  volume = {6},
  issue = {3},
  pages = {033104},
  numpages = {9},
  year = {2024},
  publisher = {American Physical Society},
  doi = {10.1103/PhysRevResearch.6.033104},
  url = {https://link.aps.org/doi/10.1103/PhysRevResearch.6.033104}
}

@article{Lin2024,
  title={{AI-Enabled Parallel Assembly of Thousands of Defect-Free Neutral Atom Arrays}}, 
  author = {{Lin, R. \emph{et al.}}},
  journal = {Phys. Rev. Lett.},
  volume = {135},
  issue = {6},
  pages = {060602},
  numpages = {7},
  year = {2025},
  month = {Aug},
  publisher = {American Physical Society},
  doi = {10.1103/2ym8-vs82},
  url = {https://link.aps.org/doi/10.1103/2ym8-vs82}
}

@article{Manetsch2024,
  author = {{Manetsch, H. J. \emph{et al.}}},
  title={{A tweezer array with 6,100 highly coherent atomic qubits
  }},
  journal = {Nature},
  year = {2025},
  volume = {647},
  pages = {60-67},
  doi = {10.1038/s41586-025-09641-4},
  url = {https://doi.org/10.1038/s41586-025-09641-4}
}

@article{Hu2024,
  title = {{Site-Selective Cavity Readout and Classical Error Correction of a 5-Bit Atomic Register}},
  author = {{Hu, B. \emph{et al.}}},
  journal = {Phys. Rev. Lett.},
  volume = {134},
  issue = {12},
  pages = {120801},
  numpages = {7},
  year = {2025},
  month = {Mar},
  publisher = {American Physical Society},
  doi = {10.1103/PhysRevLett.134.120801},
  url = {https://link.aps.org/doi/10.1103/PhysRevLett.134.120801}
}

@article{Chen_2022,
author = {{Chen, C.-C. \emph{et al.}}},
title={{Continuous Bose--Einstein condensation}},
journal={Nature},
year={2022},
month={Jun},
day={01},
volume={606},
number={7915},
pages={683-687},
issn={1476-4687},
doi={10.1038/s41586-022-04731-z},
url={https://doi.org/10.1038/s41586-022-04731-z}
}

@article{Rosi_2018,
author = {{Rosi, S. \emph{et al.}}},
title={{$\mathrm{\Lambda}$-enhanced grey molasses on the D2 transition of Rubidium-87 atoms}},
journal={Scientific Reports},
year={2018},
month={Jan},
day={22},
volume={8},
number={1},
pages={1301},
issn={2045-2322},
doi={10.1038/s41598-018-19814-z},
url={https://doi.org/10.1038/s41598-018-19814-z}
}

@article{Bernien_2022,
  title = {{Dual-Element, Two-Dimensional Atom Array with Continuous-Mode Operation}},
  author = {Singh, Kevin and Anand, Shraddha and Pocklington, Andrew and Kemp, Jordan T. and Bernien, Hannes},
  journal = {Phys. Rev. X},
  volume = {12},
  issue = {1},
  pages = {011040},
  numpages = {11},
  year = {2022},
  month = {Mar},
  publisher = {American Physical Society},
  doi = {10.1103/PhysRevX.12.011040},
  url = {https://link.aps.org/doi/10.1103/PhysRevX.12.011040}
}

@article{Thompson_2025,
  title = {{Continuous Collective Strong Coupling of Strontium Atoms to a High Finesse Ring Cavity}},
  author = {{Cline, J. R. K. \emph{et al.}}},
  journal = {Phys. Rev. Lett.},
  volume = {134},
  issue = {1},
  pages = {013403},
  numpages = {6},
  year = {2025},
  month = {Jan},
  publisher = {American Physical Society},
  doi = {10.1103/PhysRevLett.134.013403},
  url = {https://link.aps.org/doi/10.1103/PhysRevLett.134.013403}
}

@article{Greiner_2025,
author = {{Xu, M. \emph{et al.}}},
title={{A neutral-atom Hubbard quantum simulator in the cryogenic regime}},
journal={Nature},
year={2025},
month={Jun},
day={11},
issn={1476-4687},
url={https://doi.org/10.1038/s41586-025-09112-w}
}

@article{Katori_2024,
  title = {Continuous generation of an ultracold atomic beam using crossed moving optical lattices},
  author = {Okaba, Shoichi and Takeuchi, Ryoto and Tsuji, Shigenori and Katori, Hidetoshi},
  journal = {Phys. Rev. Appl.},
  volume = {21},
  issue = {3},
  pages = {034006},
  numpages = {9},
  year = {2024},
  month = {Mar},
  publisher = {American Physical Society},
  doi = {10.1103/PhysRevApplied.21.034006},
  url = {https://link.aps.org/doi/10.1103/PhysRevApplied.21.034006}
}

@article{Comparat2006,
  title = {{Optimized production of large Bose-Einstein condensates}},
  author = {{Comparat, D. \emph{et al.}}},
  journal = {Phys. Rev. A},
  volume = {73},
  issue = {4},
  pages = {043410},
  numpages = {14},
  year = {2006},
  month = {Apr},
  publisher = {American Physical Society},
  doi = {10.1103/PhysRevA.73.043410},
  url = {https://link.aps.org/doi/10.1103/PhysRevA.73.043410}
}

@article{Kurtsiefer2024,
  title = {Fano resonance in excitation spectroscopy and cooling of an optically trapped single atom},
  author = {Chow, Chang Hoong and Ng, Boon Long and Prakash, Vindhiya and Kurtsiefer, Christian},
  journal = {Phys. Rev. Res.},
  volume = {6},
  issue = {2},
  pages = {023154},
  numpages = {8},
  year = {2024},
  month = {May},
  publisher = {American Physical Society},
  doi = {10.1103/PhysRevResearch.6.023154},
  url = {https://link.aps.org/doi/10.1103/PhysRevResearch.6.023154}
}

@article{Wieman2000,
  title = {Loading an optical dipole trap},
  author = {Kuppens, S. J. M. and Corwin, K. L. and Miller, K. W. and Chupp, T. E. and Wieman, C. E.},
  journal = {Phys. Rev. A},
  volume = {62},
  issue = {1},
  pages = {013406},
  numpages = {13},
  year = {2000},
  month = {Jun},
  publisher = {American Physical Society},
  doi = {10.1103/PhysRevA.62.013406},
  url = {https://link.aps.org/doi/10.1103/PhysRevA.62.013406}
}

@article{Klosterman2022,
  title = {Fast long-distance transport of cold cesium atoms},
author = {{Klostermann, T. \emph{et al.}}},
  journal = {Phys. Rev. A},
  volume = {105},
  issue = {4},
  pages = {043319},
  numpages = {11},
  year = {2022},
  month = {Apr},
  publisher = {American Physical Society},
  doi = {10.1103/PhysRevA.105.043319},
  url = {https://link.aps.org/doi/10.1103/PhysRevA.105.043319}
}

@article{Matthies2024,
  title = {Long-distance optical-conveyor-belt transport of ultracold $^{133}\mathrm{Cs}$ and $^{87}\mathrm{Rb}$ atoms},
  author = {{Matthies, A. J. \emph{et al.}}},
  journal = {Phys. Rev. A},
  volume = {109},
  issue = {2},
  pages = {023321},
  numpages = {13},
  year = {2024},
  month = {Feb},
  publisher = {American Physical Society},
  doi = {10.1103/PhysRevA.109.023321},
  url = {https://link.aps.org/doi/10.1103/PhysRevA.109.023321}
}

@article{Schioppo2017,
author = {{Schioppo, M. \emph{et al.}}},
	da = {2017/01/01},
	doi = {10.1038/nphoton.2016.231},
	id = {Schioppo2017},
	isbn = {1749-4893},
	journal = {Nature Photonics},
	number = {1},
	pages = {48--52},
	title = {Ultrastable optical clock with two cold-atom ensembles},
	ty = {JOUR},
	url = {https://doi.org/10.1038/nphoton.2016.231},
	volume = {11},
	year = {2017}
}

@article{Biedermann2013,
  title = {{Zero-Dead-Time Operation of Interleaved Atomic Clocks}},
author = {{Biedermann, G. W. \emph{et al.}}},
  journal = {Phys. Rev. Lett.},
  volume = {111},
  issue = {17},
  pages = {170802},
  numpages = {4},
  year = {2013},
  month = {Oct},
  publisher = {American Physical Society},
  doi = {10.1103/PhysRevLett.111.170802},
  url = {https://link.aps.org/doi/10.1103/PhysRevLett.111.170802}
}

@article{Trisnadi2022,
    author = {Trisnadi, Jonathan and Zhang, Mingjiamei and Weiss, Lauren and Chin, Cheng},
    title = {Design and construction of a quantum matter synthesizer},
    journal = {Review of Scientific Instruments},
    volume = {93},
    number = {8},
    pages = {083203},
    year = {2022},
    month = {08},
    issn = {0034-6748},
    doi = {10.1063/5.0100088},
    url = {https://doi.org/10.1063/5.0100088},
}

@article{Neuhaus2024,
    author = {{Neuhaus, L. \emph{et al.}}},
    title = {{Python Red Pitaya Lockbox (PyRPL): An open source software package for digital feedback control in quantum optics experiments}},
    journal = {Review of Scientific Instruments},
    volume = {95},
    number = {3},
    pages = {033003},
    year = {2024},
    month = {03},
    issn = {0034-6748},
    doi = {10.1063/5.0178481},
    url = {https://doi.org/10.1063/5.0178481},
}

@article{Kim19,
author = {{Kim, D. \emph{et al.}}},
journal = {Opt. Lett.},
number = {12},
pages = {3178--3181},
publisher = {Optica Publishing Group},
title = {Large-scale uniform optical focus array generation with a phase spatial light modulator},
volume = {44},
month = {Jun},
year = {2019},
url = {https://opg.optica.org/ol/abstract.cfm?URI=ol-44-12-3178},
doi = {10.1364/OL.44.003178},
}

@article{KimAhn:19,
author = {Hyosub Kim and Minhyuk Kim and Woojun Lee and Jaewook Ahn},
journal = {Opt. Express},
number = {3},
pages = {2184--2196},
publisher = {Optica Publishing Group},
title = {{Gerchberg-Saxton algorithm for fast and efficient atom rearrangement in optical tweezer traps}},
volume = {27},
month = {Feb},
year = {2019},
url = {https://opg.optica.org/oe/abstract.cfm?URI=oe-27-3-2184},
doi = {10.1364/OE.27.002184},
}

@article{Ebadi2021,
author = {{Ebadi, S. \emph{et al.}}},
title={Quantum phases of matter on a 256-atom programmable quantum simulator},
journal={Nature},
year={2021},
month={Jul},
day={01},
volume={595},
number={7866},
pages={227-232},
issn={1476-4687},
doi={10.1038/s41586-021-03582-4},
url={https://doi.org/10.1038/s41586-021-03582-4}
}

@Article{Browaeys2020,
author={Browaeys, Antoine
and Lahaye, Thierry},
title={{Many-body physics with individually controlled Rydberg atoms}},
journal={Nature Physics},
year={2020},
month={Feb},
day={01},
volume={16},
number={2},
pages={132-142},
issn={1745-2481},
doi={10.1038/s41567-019-0733-z},
url={https://doi.org/10.1038/s41567-019-0733-z}
}

@article{Evered2023,
author = {{Evered, S. J. \emph{et al.}}},
title={High-fidelity parallel entangling gates on a neutral-atom quantum computer},
journal={Nature},
year={2023},
month={Oct},
day={01},
volume={622},
number={7982},
pages={268-272},
issn={1476-4687},
doi={10.1038/s41586-023-06481-y},
url={https://doi.org/10.1038/s41586-023-06481-y}
}

@article{atomcomputing2024,
  title = {{High-Fidelity Universal Gates in the ${}^{171}$$\mathrm{Yb}$ Ground-State Nuclear-Spin Qubit}},
author = {{Muniz, J. A. \emph{et al.}}},
  journal = {PRX Quantum},
  volume = {6},
  issue = {2},
  pages = {020334},
  numpages = {18},
  year = {2025},
  month = {May},
  publisher = {American Physical Society},
  doi = {10.1103/PRXQuantum.6.020334},
  url = {https://link.aps.org/doi/10.1103/PRXQuantum.6.020334}
}

@article{Endres2025,
  title = {{Benchmarking and Fidelity Response Theory of High-Fidelity Rydberg Entangling Gates}},
  author = {Tsai, Richard Bing-Shiun and Sun, Xiangkai and Shaw, Adam L. and Finkelstein, Ran and Endres, Manuel},
  journal = {PRX Quantum},
  volume = {6},
  issue = {1},
  pages = {010331},
  numpages = {28},
  year = {2025},
  month = {Feb},
  publisher = {American Physical Society},
  doi = {10.1103/PRXQuantum.6.010331},
  url = {https://link.aps.org/doi/10.1103/PRXQuantum.6.010331}
}

@article{Walhout_92,
author = {M. Walhout and J. Dalibard and S. L. Rolston and W. D. Phillips},
journal = {J. Opt. Soc. Am. B},
number = {11},
pages = {1997--2007},
publisher = {Optica Publishing Group},
title = {{$\sigma$$+$--$\sigma${\textminus} Optical molasses in a longitudinal magnetic field}},
volume = {9},
month = {Nov},
year = {1992},
url = {https://opg.optica.org/josab/abstract.cfm?URI=josab-9-11-1997},
doi = {10.1364/JOSAB.9.001997}
}

@article{JThompson2025,
  title = {{Spectroscopy and Modeling of $^{171}\mathrm{Yb}$ Rydberg States for High-Fidelity Two-Qubit Gates}},
  author = {{Peper, M. \emph{et al.}}},
  journal = {Phys. Rev. X},
  volume = {15},
  issue = {1},
  pages = {011009},
  numpages = {30},
  year = {2025},
  month = {Jan},
  publisher = {American Physical Society},
  doi = {10.1103/PhysRevX.15.011009},
  url = {https://link.aps.org/doi/10.1103/PhysRevX.15.011009}
}

@article{saffman2025,
  title = {{Universal Neutral-Atom Quantum Computer with Individual Optical Addressing and Nondestructive Readout}},
  author = {{Radnaev, A. G. \emph{et al.}}},
  journal = {PRX Quantum},
  volume = {6},
  issue = {3},
  pages = {030334},
  numpages = {20},
  year = {2025},
  month = {Aug},
  publisher = {American Physical Society},
  doi = {10.1103/66s8-jj18},
  url = {https://link.aps.org/doi/10.1103/66s8-jj18}
}

@article{Gross2017,
author = {Christian Gross  and Immanuel Bloch },
title = {Quantum simulations with ultracold atoms in optical lattices},
journal = {Science},
volume = {357},
number = {6355},
pages = {995-1001},
year = {2017},
doi = {10.1126/science.aal3837},
URL = {https://www.science.org/doi/abs/10.1126/science.aal3837}}

@article{Grangier2002,
  title = {{Collisional Blockade in Microscopic Optical Dipole Traps}},
  author = {Schlosser, N. and Reymond, G. and Grangier, P.},
  journal = {Phys. Rev. Lett.},
  volume = {89},
  issue = {2},
  pages = {023005},
  numpages = {4},
  year = {2002},
  month = {Jun},
  publisher = {American Physical Society},
  doi = {10.1103/PhysRevLett.89.023005},
  url = {https://link.aps.org/doi/10.1103/PhysRevLett.89.023005}
}

@article{Savoie2018,
author = {{Savoie, D. \emph{et al.}}},
title = {Interleaved atom interferometry for high-sensitivity inertial measurements},
journal = {Science Advances},
volume = {4},
number = {12},
year = {2018},
month = {Dec},
doi = {10.1126/sciadv.aau7948},
URL = {https://www.science.org/doi/abs/10.1126/sciadv.aau7948}}

@article{Levine2022,
  title = {{Dispersive optical systems for scalable Raman driving of hyperfine qubits}},
  author = {{Levine, H. \emph{et al.}}},
  journal = {Phys. Rev. A},
  volume = {105},
  issue = {3},
  pages = {032618},
  numpages = {13},
  year = {2022},
  month = {Mar},
  publisher = {American Physical Society},
  doi = {10.1103/PhysRevA.105.032618},
  url = {https://link.aps.org/doi/10.1103/PhysRevA.105.032618}
}

@article{holman2024trapping,
      title={{Trapping of Single Atoms in Metasurface Optical Tweezer Arrays}}, 
      author = {Holman, Aaron and \textit{et al.}},
      year={2024},
      eprint={2411.05321},
      journal={arXiv},
      primaryClass={physics.atom-ph},
      url={https://arxiv.org/abs/2411.05321}, 
}

@article{Xu2024,
author = {{Xu, Q. \emph{et al.}}},
title={Constant-overhead fault-tolerant quantum computation with reconfigurable atom arrays},
journal={Nature Physics},
year={2024},
month={Jul},
day={01},
volume={20},
number={7},
pages={1084-1090},
issn={1745-2481},
doi={10.1038/s41567-024-02479-z},
url={https://doi.org/10.1038/s41567-024-02479-z}
}

@article{Baranes2025,
    title={{Leveraging Atom Loss Errors in Fault Tolerant Quantum Algorithms}}, 
    author = {{Baranes, G. \emph{et al.}}},
    year={2025},
    eprint={2502.20558},
    journal={arXiv},
    primaryClass={quant-ph},
    url={https://arxiv.org/abs/2502.20558} 
}

@article{Wang2023,
  title = {{Accelerating the Assembly of Defect-Free Atomic Arrays with Maximum Parallelisms}},
  author = {{Wang, S. \emph{et al.}}},
  journal = {Phys. Rev. Appl.},
  volume = {19},
  issue = {5},
  pages = {054032},
  numpages = {10},
  year = {2023},
  month = {May},
  publisher = {American Physical Society},
  doi = {10.1103/PhysRevApplied.19.054032},
  url = {https://link.aps.org/doi/10.1103/PhysRevApplied.19.054032}
}

@article{atomcomputing2025,
      title={Repeated ancilla reuse for logical computation on a neutral atom quantum computer}, 
      author = {{Muniz, J. A. \emph{et al.}}},
      year={2025},
      eprint={2506.09936},
      journal={arXiv},
      primaryClass={quant-ph},
      url={https://arxiv.org/abs/2506.09936}, 
}

@article{Li2025,
      title={Fast, continuous and coherent atom replacement in a neutral atom qubit array}, 
      author = {{Li, Y. \emph{et al.}}},
      year={2025},
      eprint={2506.15633},
      journal={arXiv},
      primaryClass={quant-ph},
      url={https://arxiv.org/abs/2506.15633}, 
}

@article{Breuckmann2021,
  title = {{Quantum Low-Density Parity-Check Codes}},
  author = {Breuckmann, Nikolas P. and Eberhardt, Jens Niklas},
  journal = {PRX Quantum},
  volume = {2},
  issue = {4},
  pages = {040101},
  numpages = {19},
  year = {2021},
  month = {Oct},
  publisher = {American Physical Society},
  doi = {10.1103/PRXQuantum.2.040101},
  url = {https://link.aps.org/doi/10.1103/PRXQuantum.2.040101}
}

@article{Bravyi2024,
author = {{Bravyi, S. \emph{et al.}}},
title={High-threshold and low-overhead fault-tolerant quantum memory},
journal={Nature},
year={2024},
month={Mar},
day={01},
volume={627},
number={8005},
pages={778-782},
issn={1476-4687},
doi={10.1038/s41586-024-07107-7},
url={https://doi.org/10.1038/s41586-024-07107-7}
}

@article{tomita2024,
      title={Atom Camera: Super-resolution scanning microscope of a light pattern with a single ultracold atom}, 
      author = {{Tomita, T. \emph{et al.}}},
      year={2024},
      eprint={2410.03241},
      journal={arXiv},
      primaryClass={physics.atom-ph},
      url={https://arxiv.org/abs/2410.03241}, 
}

\setcounter{figure}{0}
\newcounter{EDfig}
\renewcommand{\figurename}{Extended Data Fig.}

\begin{extendedfigure*}[p]
\centering
\includegraphics[width=1\textwidth]{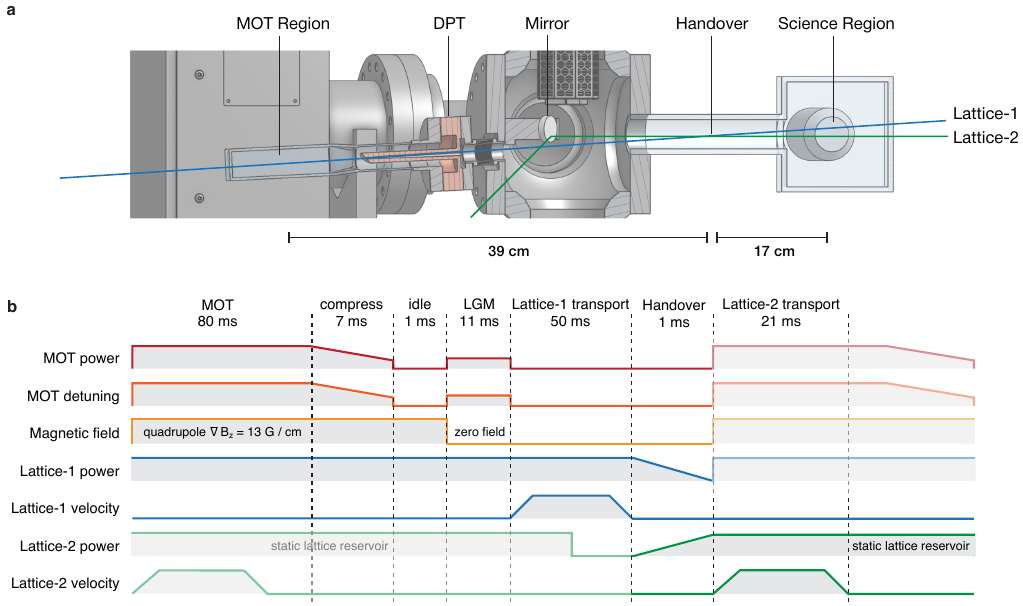}
    \caption{\justifying
    \textbf{Vacuum chamber and dual-lattice transport sequence.}
    \textbf{a,} Simplified view of the vacuum chamber. Atoms are cooled and loaded from a MOT into an optical lattice (Lattice-1) and then transported through the differential pumping tube (DPT, orange) to the science chamber. The atomic cloud is then handed over to a second optical lattice (Lattice-2), which is reflected out of the chamber by an in-vacuum mirror. While most MOT light is blocked by the DPT, the tilted design between both chambers avoids direct line-of-sight between the computational array and the MOT location.
    \textbf{b,} Summary of the dual-lattice transport sequence, including the MOT stage, loading and cooling of Lattice-1 as well as lattice transport and handover. After the lattice handover, we restart the lattice loading procedure in the MOT chamber, while atoms in Lattice-2 are shipped to the science region where they serve as an atomic reservoir for tweezer extraction. In the Lattice-2 velocity graph, the brief back-and-forth movement used to avoid atom spilling during qubit preparation is omitted for clarity. The gray-shaded regions indicate the previous/next lattice loading cycle.
    }
\label{fig:suppfig_vac}
\end{extendedfigure*}

\begin{extendedfigure*}[p]
\centering
\includegraphics[width=1\textwidth]{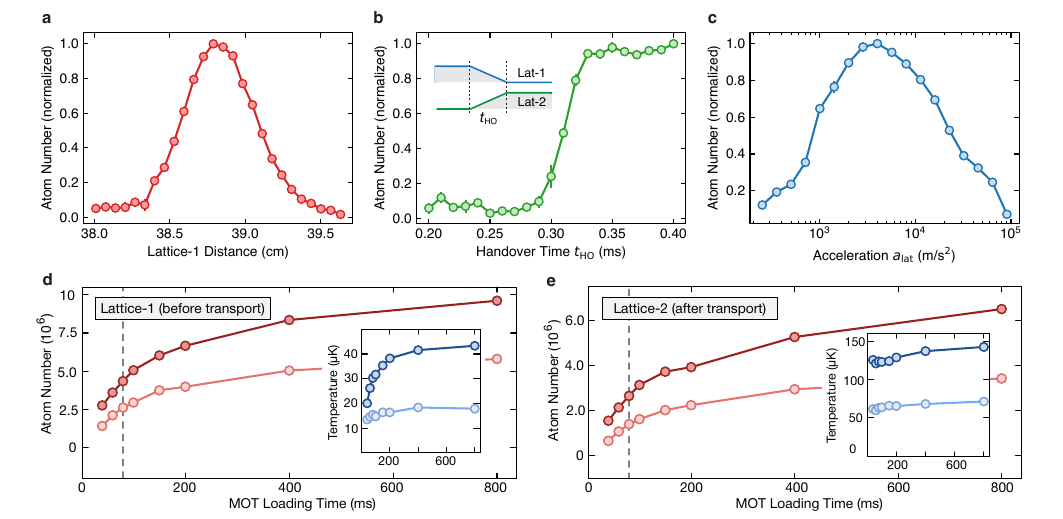}
    \caption{\justifying
    \textbf{Dual-lattice conveyor belt transport.}
    \textbf{a,} Atom number (normalized) obtained in the reservoir after transport as a function of Lattice-1 travel distance before handing the atomic cloud over to Lattice-2.
    \textbf{b,} Atom number (normalized) obtained in the reservoir as a function of time dedicated for lattice handover, which consists of a simultaneous and opposite ramp of the Lattice-1 and Lattice-2 intensities (see inset schematic). 
    \textbf{c,} Atom number (normalized) obtained in the reservoir as a function of conveyor belt acceleration, here shown exemplary for Lattice-1.
    \textbf{d, e,} Atom number (red) and temperature (blue) in the respective transport lattices before (d) and after (e) transport, shown for varying MOT loading times with the lattices $\sim\!300\,$GHz red-detuned from the D$_1$ line (darker color shading). The dashed gray line represents our chosen MOT loading time, which was sufficient to obtain the reservoir density required for tweezer loading. The light-shaded curves are measured for further red-detuned lattices ($\sim\!700\,$GHz), providing lower lattice-induced scattering and colder temperatures but also lower atom numbers. Error bars represent the standard error of the mean across 10 repetitions.}
\label{fig:suppfig_lattice}
\end{extendedfigure*}

\begin{extendedfigure*}[p]
\centering
\includegraphics[width=\textwidth]{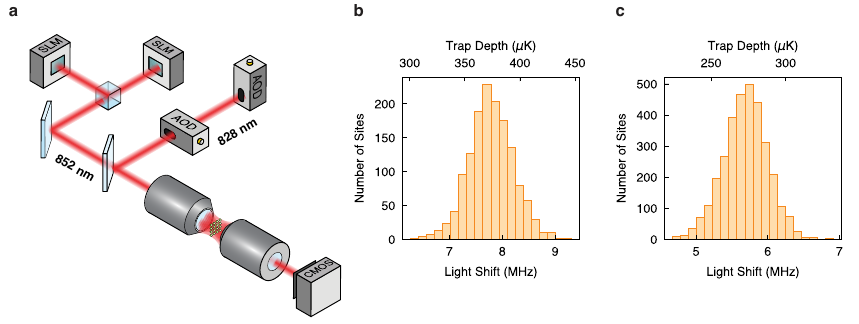}
    \caption{\justifying
    \textbf{High-NA beam paths and optical tweezer characterization.}
    \textbf{a,} Simplified optical beam paths for tweezer generation and single-atom imaging. Before the tweezer-generation objective, the preparation and storage zone SLMs (both at 852\,nm) are combined on a PBS, then further combined with the 828\,nm AOD beam path on a dichroic beamsplitter. A second objective is used for single-atom imaging. Relay optics are omitted for simplicity.
    \textbf{b, c,} Histograms of trap depth measured for the preparation zone (b) and storage zone (c) tweezer arrays after two-stage homogenization. The preparation (storage) zone tweezers have an average trap depth of 370\,$\upmu\mathrm{K}$ (270\,$\upmu\mathrm{K}$) with standard deviation 5.4\% each.
    }
\label{fig:suppfig_tweezers}
\end{extendedfigure*}

\begin{extendedfigure*}[p]
\centering
\includegraphics[width=1\textwidth]{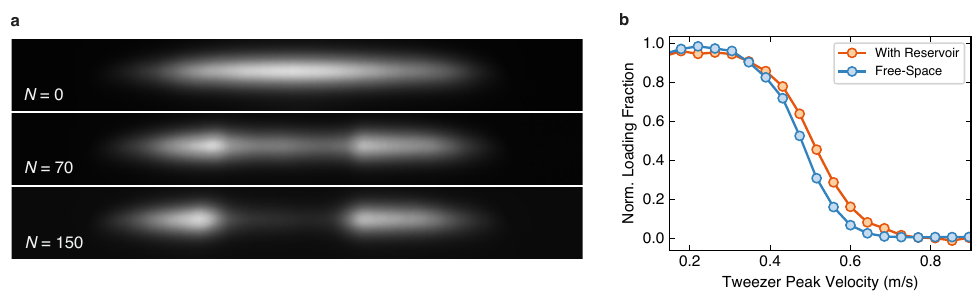}
    \caption{\justifying
    \textbf{Extracting atoms from the lattice reservoir.}
    \textbf{a,} Single-shot fluorescence images of the lattice reservoir after extracting atoms via optical tweezers for $N$-repetitions.
    \textbf{b,} Normalized tweezer array loading fraction obtained in the preparation zone as a function of tweezer velocity during extraction from the reservoir. We find no significant difference when moving atoms perpendicularly through the lattice potential versus in free-space, with a slightly higher survival possibly attributed to the superimposed lattice-tweezer potential during atom transport. Error bars represent the standard error of the mean across 10 repetitions.
    }
\label{fig:suppfig_twzloading}
\end{extendedfigure*}

\begin{extendedfigure*}[p]
\centering
\includegraphics[width=1\textwidth]{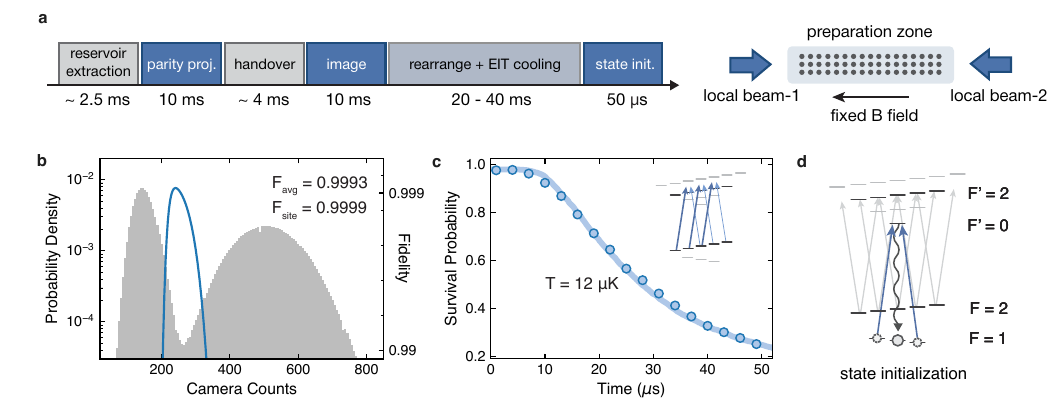}
    \caption{\justifying
    \textbf{Characterization of the qubit preparation sequence.}
    \textbf{a,} Summary of the experimental sequence for qubit preparation. After extraction from the reservoir, the atoms are transported to the preparation zone within $\sim\!2.5\,\mathrm{ms}$. Here, a parity-projection pulse of $10\,\mathrm{ms}$ is performed to achieve either one or zero atoms per AOD tweezer trap. Atoms are then handed off to a backbone tweezer array generated by an SLM while applying PGC, involving SLM-AOD intensity ramps and a brief ($30\,\upmu\mathrm{s}$) resonant Talbot plane push-out pulse. A $10\,\mathrm{ms}$ fluorescence image is used to identify occupied sites for rearrangement, followed by EIT cooling and simultaneous atom sorting into a defect-free array. Depending on the camera settings and desired atom configuration, rearrangement takes between $20\,\mathrm{ms}$ and $40\,\mathrm{ms}$ (largely dominated by data transfer latency). Finally, atoms are optically pumped into the qubit state $\ket{0}$ within $50\,\upmu\mathrm{s}$. All light pulses are performed by two circularly-polarized counter-propagating laser beams at static magnetic field (schematic).
    \textbf{b,} Imaging histogram for 1D imaging at finite magnetic field. The extracted average discriminant fidelity is 0.9993 with a site-resolved discriminant fidelity of 0.9999. The blue curve visualizes the discriminant fidelity as a function of threshold value. Note that the net imaging fidelity must also account for imaging survival, which limits total fidelity to $\mathcal{F} \sim 0.995$.
    \textbf{c,} Drop-and-recapture measurement of atomic temperature after $40\,\mathrm{ms}$ of 1D EIT cooling and subsequent qubit state preparation. The temperature is extracted as $T = 12\,\upmu\mathrm{K}$ via Monte-Carlo simulations. Inset: Relevant atomic levels for EIT cooling.
    \textbf{d,} Schematic illustrating relevant atomic levels for fast initialization of qubit state $\ket{0}$. Error bars represent the standard error of the mean across 10 repetitions.
    }
    \label{fig:suppfig_stateprep}
\end{extendedfigure*}

\begin{extendedfigure*}[p]
\centering
\includegraphics[width=\textwidth]{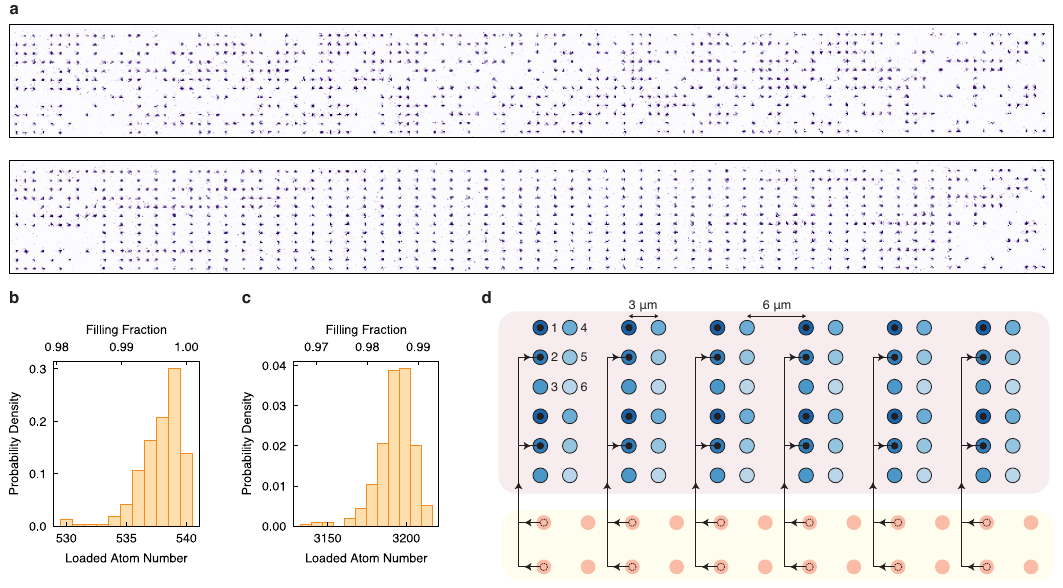}
    \caption{\justifying
    \textbf{Preparation zone atom rearrangement and iterative assembly of the storage zone array.}
    \textbf{a,} Single-shot fluorescence images of the preparation zone atom array. Before rearrangement, atoms are stochastically loaded (top). After rearrangement, the target array is filled with near-unity probability (bottom). For this shot, the 600-site target array has zero defects. Explicitly ejecting atoms outside the target array is not necessary, as only the target array is picked up by AOD tweezers and transported to the storage zone. \textbf{b,} Preparation zone rearrangement histogram (500 trials) of 540 target sites. We achieve a 99.6\%  average filling fraction with 14\% of trials having zero defects. For visualization, trials with fewer than 530 atoms ($\approx\!1\%$ of all trials) are grouped into the 530-atom bin. \textbf{c,}  Storage zone loading histogram (300 trials). We achieve an average loading fraction of 98.5\% (3{,}193 atoms). \textbf{d,} Storage zone iterative assembly scheme. In each iteration, 540 AOD tweezers pick up sorted atoms from the preparation zone target array (bottom), which are transported to the storage zone (top) to sequentially fill one of six subarrays (blue color shading). The AOD tweezers travel through wide channels in the storage array to avoid crosstalk with the static SLM tweezer traps. The schematic illustrates filling of the second subarray, where the first subarray has already been filled. See also Supplementary Video 1.
    }
\label{fig:suppfig_rearrange}
\end{extendedfigure*}

\begin{extendedfigure*}[p]
\centering
\includegraphics[width=1\textwidth]{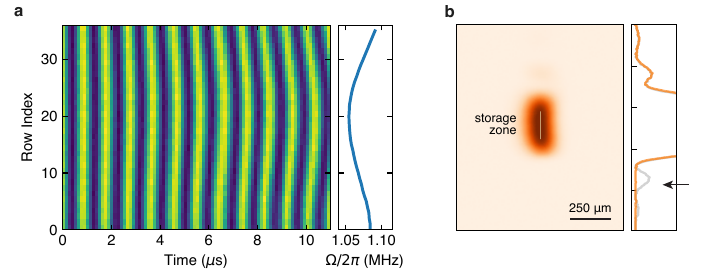}
    \caption{\justifying
    \textbf{Raman and shielding beam characterization.}
    \textbf{a,} Rabi oscillations between qubit states for each row of the storage array. The variation of fitted Rabi frequency allows us to extract a Raman beam homogeneity of approximately $1.04\%$ root-mean-square variation and $3.4\%$ peak-to-peak variation on the atoms.
    \textbf{b,} Beam profile of the knife-edged flat-top 1529\,nm shielding light, where the storage zone location is indicated. The plot on the right shows a zoomed-in line profile through the center of the beam, where the orange (gray) curve corresponds to the profile with (without) the knife edge. Evidently, cropping residual beam tails is imperative to avoid beam cross talk into the preparation zone (location marked with the black arrow).
    }
\label{fig:suppfig_raman1530}
\end{extendedfigure*}

\begin{extendedfigure*}[p]
\centering
\includegraphics[width=1\textwidth]{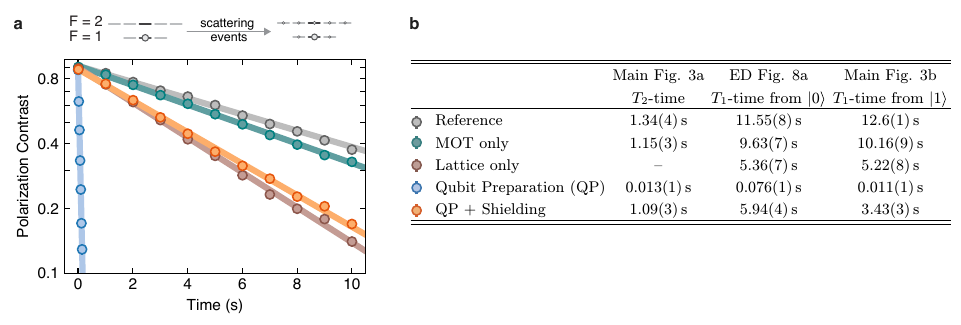}
    \caption{\justifying
    \textbf{Qubit depolarization under various conditions.}
    \textbf{a,} In main text Fig.\,\ref{fig:fig4}b, we investigate how various parallel operations influence qubit polarization when initialized in state $\ket{1}$, where we expect a strong effect as pumping lights used for MOT and qubit preparation predominantly operate from $F=2$. Here, we show a complementary analysis for qubits initialized in $\ket{0}$. Similarly comparing storage qubit depolarization during local qubit preparation with and without shielding, we are able to recover the measured $T_1$-time up to the depolarization rate set by the lattice lights. As expected, we generally find lower depolarization rates compared to starting in $\ket{1}$.
    \textbf{b,} Summary of the fitted $1/e$ $T_2$- and $T_1$-times from Figs.\,\ref{fig:fig4}a,b and Extended Data Fig.\,8a. Errors and error bars represent the standard error of the mean across 10 repetitions.
    }
\label{fig:supp_t1t2}
\end{extendedfigure*}

\begin{extendedfigure*}[p]
\centering
\includegraphics[width=1\textwidth]{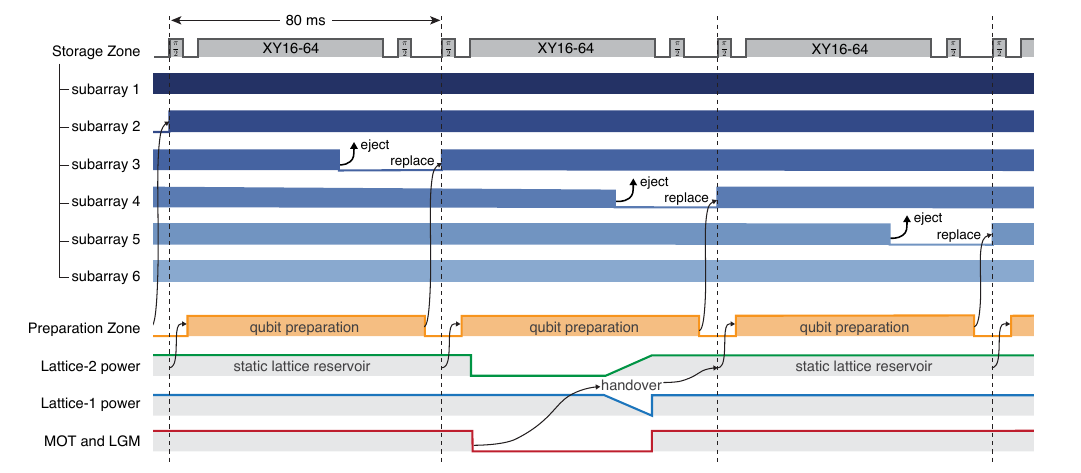}
    \caption{\justifying
    \textbf{Experimental sequence for continuous operation while maintaining qubit coherence.} Summary of the sequence used to achieve the results of Figs.\,\ref{fig:fig3}b,c,d and Extended Data Figs.\,\ref{fig:quantumbattery}a,b. In the storage zone, dynamical decoupling (except for Fig.\,\ref{fig:fig3}b) and shielding are continuously applied to the storage qubits, while the oldest qubit subarray is discarded and refilled with fresh qubits from the reservoir. First, an initial $\text{X}_{\pi/2}$ pulse prepares qubits in a coherent superposition state. After XY16-64 decoupling, (remaining) coherence is briefly mapped back to population, typically into $\ket{0}$ with a $\text{X}_{-\pi/2}$ pulse, before the next qubit subarray is introduced. Thereby, qubits are in the equal superposition state for about 90\% of total experiment duration. In Extended Data Fig.\,\ref{fig:quantumbattery}b, we show an example of mapping back into alternating $\ket{0}$ and $\ket{1}$ populations by using a final $\text{X}_{+\pi/2}$ pulse instead. Throughout the experimental sequence, laser cooling in the MOT chamber, dual-lattice transport, and qubit preparation run in parallel in the background to provide a high-rate qubit supply. For Fig.\,\ref{fig:fig3}b, no decoupling pulses are applied to the storage zone and we simply probe qubit polarization.
    }
\label{fig:contcohseq}
\end{extendedfigure*}

\begin{extendedfigure*}[p]
\centering
\includegraphics[width=1\textwidth]{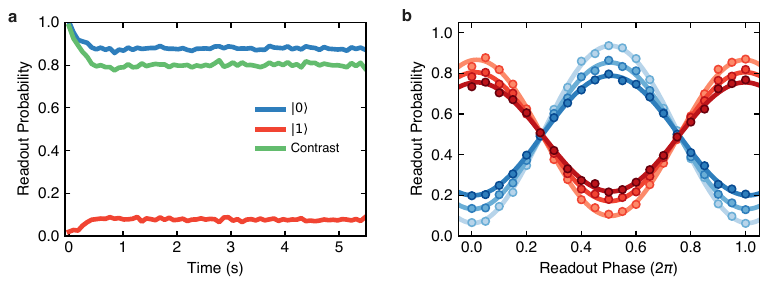}
    \caption{\justifying
    \textbf{Coherence under continuous operation.}
    \textbf{a,} Complementary analysis to Fig.\,\ref{fig:fig3}c when array-averaging the probability to read out qubits in state $\ket{0}$ (blue) or $\ket{1}$ (red). The green line indicates the contrast, i.e. the difference of populations measured in $\ket{0}$ and $\ket{1}$.
    \textbf{b,} Similar to Fig.\,\ref{fig:fig3}d, but instead with the decoupling sequence $(\text{X}_{\pi/2}-\text{XY16-64}-\text{X}_{+\pi/2})$ applied during each reloading cycle. This results in alternating qubit states in the storage array (checkerboard pattern). The blue (red) curves represent the readout probability of even (odd) subarrays, indicated by the color shading. Errors and error bars represent the standard error of the mean across 10 repetitions.
}
\label{fig:quantumbattery}
\end{extendedfigure*}

\newpage

\section*{Methods}

\noindent\textbf{Vacuum System} \\ 
A simplified schematic of our vacuum system is shown in Extended Data Fig.\,\ref{fig:suppfig_vac}a. The system consists of a MOT chamber and a science chamber, separated by a custom-designed differential pumping tube (DPT, Limit Vacuum Technology) with a 1.5\,mm back aperture and 4.3\,mm front aperture. The DPT maintains a pressure differential between the two chambers and blocks most of the MOT light. Both DPT and MOT chamber are tilted by approximately $4^\circ$ to prevent direct scattering of cooling light onto the atomic array in the science chamber, where the line-of-sight passes about 1\,cm above the array location. The MOT chamber is primarily composed of a glass cell (Precision Glassblowing) with two rubidium dispenser arms (not shown in Extended Data Fig.\,\ref{fig:suppfig_vac}). The science chamber features a double-sided anti-reflection-coated glass cell (Akatsuki Technology Co. Ltd) with optical contact technology. In both chambers, the pressure remains below the measurable threshold of the ion pumps (SAES NEXTorr and Agilent StarCell 75). Several components are omitted from the figure for clarity, including in-vacuum electrodes (not used in this work) and a vacuum viewport, which provides optical access to the in-vacuum mirror.\\

\noindent\textbf{Objective \& Imaging System} \\ 
The experimental setup features a high-NA optical system, which enables high-efficiency imaging and tight trapping of single atoms over a field-of-view of more than 1.5\,mm diameter (Extended Data Fig.\,\ref{fig:suppfig_tweezers}a). At its core are two NA=0.65 objectives (Special Optics, custom-design). One of the objectives is used for projecting optical tweezers, and the other for single-atom imaging. The two objectives maintain diffraction-limited performance across the entire field-of-view for wavelengths ranging from 780 to 860\,nm. The objective's optical transmission is 92\%, and we estimate the total absorption to be $\sim\!1\%$ (taking into account finite reflection at each AR-coated surface), which reduces thermal lensing and enables higher trapping laser power in the future.

For single-atom imaging, we use two 4-f telescopes (one high-NA objective and three relay lenses) to map the atomic plane inside the glass cell onto a low-noise CMOS camera (Hamamatsu C15550-20UP). The imaging system magnification is 7.6, such that the fluorescence of a single atom is mapped onto $\approx\!3\times 3$ camera pixels. The quantum efficiency of the CMOS camera at the 780\,nm imaging wavelength is $\sim\!50\%$. All relay lenses used in the objective beam path (both for imaging and tweezer projection) are custom-designed (Special Optics) to accommodate the large field-of-view.\\

\noindent\textbf{Tweezer Generation} \\
For optical tweezer projection, we use polarizing beamsplitters (PBS) and dichroic beamsplitters to combine three separate beam paths powered by three high-power lasers: a 15\,W, 828\,nm fiber amplifier system (Precilaser) that generates the dynamic optical tweezers for atom transport, and two 15\,W, 852\,nm fiber amplifier systems (Precilaser) that each form the backbone static tweezer array in the preparation and storage zones (Extended Data Fig.\,\ref{fig:suppfig_tweezers}a).

The 828\,nm dynamic tweezers beam path, dedicated to atom transport and sorting, consists of two perpendicularly mounted AODs (G\&H AODF 4085) separated by a 1-to-1 4-f telescope. Another 4-f system maps the AOD aperture to the Fourier plane of the objective. The AOD-generated tweezers have a waist of $\approx\!800\,$nm and travel range of $600\,\upmu\mathrm{m}$ in each dimension on the atom plane. Depending on the transport pattern required, we dynamically switch between different tweezer configurations within one cycle of the experiment. When extracting atoms from the reservoir to the preparation zone, we use 1{,}440 tweezers at $4.5\,\upmu\mathrm{m}$ spacing with average depth of $450\,\upmu\mathrm{K}$. To transport sorted atoms to or eject atoms from the storage zone, we generate an array of 540 tweezers at $9\,\upmu\mathrm{m}$ spacing with an average depth of $600\,\upmu\mathrm{K}$. We empirically find a reduction in tweezer lifetime as we reduce AOD tweezer spacing, potentially due to atom heating from beating between residual optical potentials of neighboring tweezers.

Static optical tweezer arrays in the preparation and the storage zones are generated in two separate beam paths by two independent SLMs (Hamamatsu X15213-02R) and then combined on a PBS. Each beam path includes a 4-f relay lens system to map the SLM aperture to the Fourier plane of the objective. The SLM phase pattern is calculated using a variation of the Weighted Gerchberg-Saxton (WGS) algorithm\cite{Kim19}, and calculation accelerated with a GPU. We numerically ``pad" the SLM with 0's such that the 2D SLM field array (iteratively optimized using WGS algorithm) is 10$\times$10 times larger than the SLM pixel number, enabling 10 times finer control over tweezer positions\cite{KimAhn:19}. This corresponds to a tweezer positioning precision of $65\,$nm, which reflects an order of magnitude improvement over the natural diffraction unit of $650\,$nm.

We find significant tweezer spacing distortion due to non-linear effects across the large array span. To systematically overlap thousands of AOD and SLM tweezers, we run an automated procedure that images both AOD and SLM tweezers on a camera, calculates the displacement between the two sets of tweezers for each site, and feeds back on the target tweezer positions of the WGS algorithm site-by-site. We also apply Zernike polynomials to correct for aberrations in the optical system\cite{Ebadi2021}, which increases tweezer trap depth by $\sim\!10\%$ post-correction.

SLM diffraction efficiency decreases as the distance to the 0-th order increases. To homogenize our backbone tweezer arrays in the preparation and storage zone, we apply the following two-step procedure. First, when generating optical tweezer arrays using the WGS algorithm, we precompensate for spatially varying diffraction efficiency by including a ``sinc" term in the target array\cite{Ebadi2021}. This rough homogenization typically yields 15\% to 20\% inhomogeneity. Then, we run an atom-based homogenization procedure that relies on site-resolved measurements of tweezer-induced light shifts, which is used to feed back onto the WGS target intensity at each site. In the preparation zone, we measure the tweezer light shift by probing the $F = 2 \rightarrow F' = 2$ transition; in the storage zone, we infer the differential light shift via Ramsey interferometry between the two qubit states\cite{tomita2024}. After a few rounds of atom-based feedback, we arrive at 5$\sim$6\% inhomogeneity in both preparation and storage zone (Extended Data Fig.\,\ref{fig:suppfig_tweezers}b,c). After aberration correction and homogenization, the average SLM trap depth is $370\,\upmu\mathrm{K}$ ($270\,\upmu\mathrm{K}$) in the preparation (storage) zone with tweezer waist $\approx 800\,$nm.\\

\noindent\textbf{MOT \& Lattice Loading} \\
The experiment starts with the preparation of an atom reservoir, see Extended Data Fig.\,\ref{fig:suppfig_vac}b. We first load $\approx\!10^7$ atoms in a MOT within $80\,\mathrm{ms}$ while the first optical lattice conveyor belt (Lattice-1) is overlapped throughout. The MOT light is $23\,\mathrm{MHz}$ red-detuned from the hyperfine transition $F=2 \rightarrow F'=3$, where $F$ refers to hyperfine levels in the $5S_{1/2}$ ground state and $F'$ to hyperfine levels in the $5P_{3/2}$ excited state. The repumping light, created via modulating a sideband on the cooling light, resonantly drives the $F=1 \rightarrow F'=2$ transition. We operate the MOT at a magnetic field gradient of 13\,G/cm, and use a 395\,nm UV-LED for light-induced atom desorption from the glass cell. After the MOT stage, the MOT light is ramped to lower intensity and about $140\,\mathrm{MHz}$ red-detuning over 7\,ms for compression of the atomic cloud into the lattice. A brief idle time follows the lattice loading procedure, in which the cooling lights are switched off and the magnetic field is zero-ed. Subsequently, we perform Lambda-Gray Molasses (LGM)\cite{Rosi_2018} at low cooling light intensity, with the carrier frequency placed $30\,\mathrm{MHz}$ blue-detuned from the $F=2 \rightarrow F'=2$ transition and the coherent repumper sideband on two-photon resonance with the $F=1 \rightarrow F=2$ transition between both hyperfine ground states. After $t_{\mathrm{LGM}} = 11\,\mathrm{ms}$, we load $\approx\!4\times 10^6$ atoms at temperatures $T\approx 20\,\upmu\mathrm{K}$ into Lattice-1 as measured via absorption imaging.\\

\noindent\textbf{Dual-Lattice Optical Transport} \\
We transport atoms from the MOT to the science region using two angled conveyor belt optical lattices\cite{Katori_2024, Klosterman2022,Matthies2024}. Both transport lattices are derived from a single Ti:Sapphire laser (Matisse, Spectra-Physics) and $\sim\!300\,$GHz red-detuned from the D$_1$ line, which is found to be the empirical optimum for our available laser power (Extended Data Fig.\,\ref{fig:suppfig_lattice}d,e). Lattice-1 has a Gaussian beam waist of around $330\,\si{\micro\meter}$ at the position of the MOT and a minimum waist of around $250\,\si{\micro\meter}$. Particularly at the position of the DPT, its beam diameter is roughly three times smaller than the DPT aperture. Both conveyor belt lattices are spatially mode-matched at the handover-point, from which the waist of the second conveyor belt lattice (Lattice-2) decreases to a minimum waist of $150\,\si{\micro\meter}$ in the microscope field-of-view. Both conveyor belt lattices are created via retro-reflection of their respective incoming beams, which are deflected by two acousto-optical modulators (AOMs) into opposite diffraction orders, then imaged back onto the lattice waist (quad-pass configuration)\cite{Trisnadi2022}. Mounting the AOMs perpendicular to each other ensures a circular beam shape and enables optimal overlap with the incoming lattice beam on retro-reflection. The quad-pass efficiency is $(0.88)^4 \approx 60\%$, such that for typical incoming powers $P_{\mathrm{in}}\approx 1\,$W we achieve lattice depths $U_{\mathrm{lat}} > 500\,\si{\micro\kelvin}$ for both conveyor belt lattices across the entire transport distance.

After loading Lattice-1, we linearly ramp the frequency of one of the retro-AOMs to introduce a frequency detuning $\upDelta \nu(t)$ between both interfering laser beams, and obtain a conveyor belt lattice moving at velocity $v = \lambda \upDelta\nu(t)/2$, with $\lambda$ the wavelength of the lattice laser\cite{Klosterman2022,Matthies2024}. The atom cloud is transported over $\sim\!39\,\mathrm{cm}$ (Extended Data Fig.\,\ref{fig:suppfig_lattice}a) before arriving at the handover-point after $t_{\mathrm{L1}} = 50\,\mathrm{ms}$. Here, we transfer the atomic cloud from Lattice-1 to Lattice-2 within $t_{\mathrm{HO}} = 1\,\mathrm{ms}$ by a simultaneous and opposite linear intensity ramp of both lattice lights and without applying cooling light during the handover (Extended Data Fig.\,\ref{fig:suppfig_lattice}b). Finally, within $t_{\mathrm{L2}} = 21\,\mathrm{ms}$, Lattice-2 transports the atoms over another $\sim\!17\,\mathrm{cm}$ into the microscope field-of-view, where it serves as an atom reservoir. For both conveyor belt lattices, we find optimal transport efficiency for accelerations $a_{\mathrm{lat}}\approx 4{,}000\,\mathrm{m/s^2}$ (Extended Data Fig.\,\ref{fig:suppfig_lattice}c) and velocities of 8-10\,m/s, limited by AOM bandwidth. Using this scheme, we deliver reservoirs of approximately $2.5\times 10^6$ atoms at a temperature around $120\,\upmu\mathrm{K}$ into our reloading zone, which corresponds to an approximately $60\%$ dual-lattice transport efficiency at $6\times$ the original temperature. Most of the observed heating is attributed to the lattice handover, not the long-distance transport.

When periodically replacing the atomic reservoir, we start loading a new MOT directly after the lattice handover and, as such, can deliver a fresh reservoir cloud to the science region every approximately $150\,$ms. It is noted that only during the second stage of lattice transport, $t_{\mathrm{L2}} = 21\,\mathrm{ms}$, no reservoir is available for loading optical tweezers; however, as described in the main text, this is not a limitation to continuous operation as the qubit preparation procedure typically exceeds $t_{\mathrm{L2}}$.

Compared to a single transport lattice design\cite{Klosterman2022,Matthies2024}, our dual-lattice architecture offers several advantages: (1) Due to the angle between both transport lattices and the differential pumping tube aperture, we avoid a direct line-of-sight and therefore reduce scattering from the MOT onto qubits already present in the science region. (2) The waist of the second conveyor belt lattice becomes largely independent of overall transport distance, and therefore can be decreased within the field-of-view of the objective. This increases the reservoir density for tweezer loading and also limits the impact of lattice-induced scattering and dipole potentials on other zones. (3) While Lattice-2 is still in active use, atoms in Lattice-1 can be prepared and transported to the handover-point in parallel. This enables fast sequential reservoir replacement, and decouples the MOT and lattice loading sequence from science chamber operations.\\

\noindent\textbf{Tweezer Loading from the Lattice Reservoir} \\
In this work, we load optical tweezers from the dense lattice reservoir without employing additional laser cooling during the loading process. Here, we briefly discuss our current understanding of the underlying mechanisms, and outline the effect of loading and extracting tweezers from an active reservoir. In our parameter regime, we expect two mechanisms to contribute to our tweezer loading: \textit{Stochastic} and/or \textit{collisional} loading. Both critically depend on atomic density $n(r,z)$ in the reservoir, which is a function of atom number $N$ per lattice site, atom temperature $T$, and both radial and axial trapping frequencies $\omega_r$ and $\omega_z$. Within one lattice site, it is given by
\begin{equation*}
    n(r,z) = n_0 \exp{\left(-\frac{m}{2k_B T}\left( w_r^2r^2 + w_z^2z^2\right)\right)}
\end{equation*} 
with $n_0 = N \omega_r^2 \omega_z \left( m/(2 \pi k_B T)\right)^{3/2}$ the peak density, $m$ the mass of $^{87}$Rb, $r$ the radial distance from the center of the reservoir, and $z$ the axial distance from the lattice site. Neglecting dependencies on atom temperature and the relative trap depth between the lattice and tweezers, we now turn to a brief discussion of both loading mechanisms.

\noindent\textit{Stochastic loading:} When turning on the tweezer, the number of reservoir atoms stochastically overlapped with the tweezer volume can be approximated as $N_{st} \propto V_{\mathrm{twz}} \langle n(r=0) \rangle_{\mathrm{lat}}$, where we assume peak density in the radial direction and $\langle \, \rangle_{\mathrm{lat}}$ denotes averaging across the axial lattice sites. These are valid approximations, as the tweezers are far smaller than the radial dimension of the reservoir but capture multiple lattice sites axially. For our reservoir parameters, we estimate an atom density of $\langle n(r = 0) \rangle_{\mathrm{lat}}\approx 5\times 10^{11}\,\mathrm{cm}^{-3}$ and tweezer volume $V_\mathrm{twz} \approx10^{-17}\,\mathrm{m}^3$, which results in $\sim\!5$ atoms stochastically overlapped with the tweezer volume.

\noindent\textit{Collisional loading}: 
An atom moving through a background gas collides with a rate $\Gamma(r,z) = n(r,z) v_{\mathrm{rel}}\sigma$, with $v_{\mathrm{rel}}$ the thermal relative velocity between two atoms, and $\sigma$ the scattering cross section. The two-body collision density is given by $\gamma(r,z) = \frac{1}{2} n(r,z)^2 v_{\mathrm{rel}} \sigma$, such that the collision rate within the tweezer volume is proportional to $\Gamma_\mathrm{twz} \propto V_{\mathrm{twz}} \langle \gamma(r = 0) \rangle_{\mathrm{lat}}$ using the same approximations as above. For our system parameters, we estimate $\langle \gamma(r = 0) \rangle_{\mathrm{lat}} \approx 3\times10^{19}\,\mathrm{m}^{-3}\mathrm{s}^{-1}$ and therefore loading of $\sim\!0.3$ atoms per millisecond of tweezer-lattice overlap, assuming one atom trapped per atomic collision.

While these order-of-magnitude estimates imply  that stochastic loading is dominating, we would like to point out two simplistic assumptions this model is making: (a) While spatial overlap is necessary, it is not a sufficient condition for trapping; additionally considering kinetic energy constraints would lead to a decrease of stochastically captured atoms. (b) Due to the abruptly altered potential landscape on tweezer turn-on, nearby atoms accelerate toward the trap center and the atomic collision rate within the tweezer increases; this results in a larger number of atoms loaded via collisions\cite{Comparat2006}. For these reasons, we expect both mechanisms to contribute and further investigation is warranted.

In the regime of few tweezer extractions (i.e., saturated loading, see also Fig.\,1b), we find that, on average, $\approx\!5$ atoms are lost from the reservoir per tweezer extraction, which is of the same order as the combined tweezer loading estimates above. This number is measured by comparing reservoir atom number with the number of atoms obtained in tweezers from repeated extraction, accounted for parity-projection. It should be noted that this serves as an upper bound and does not necessarily imply \textit{loading} of 5 atoms per tweezer, since we expect the rather invasive extraction procedure to accelerate evaporative losses in the reservoir as well. After loading, captured atoms are transported out of the active reservoir perpendicular to the axial lattice potential. As shown in Extended Data Fig.\,4b, we observe no difference in survival when moving atoms through the reservoir versus in free-space as a function of tweezer velocity\cite{Gyger2024}.\\

\noindent\textbf{Finite-Field Laser Cooling \& Imaging} \\
A key feature of our reloading architecture is the ability to initialize fresh qubits and perform mid-circuit laser cooling and imaging in the presence of a static magnetic field. Avoiding field changes, such as field-zeroing as typically necessary for polarization gradient cooling (PGC), protects coherence in existing qubits and enables faster qubit preparation cycles. To this end, all qubit preparation and manipulation protocols are designed to operate at a fixed magnetic field of 4.2\,G, which defines the qubit quantization axis at all times. The preparation zone is illuminated with a pair of 1D counter-propagating 780\,nm beams with opposite circular polarizations ($\sigma^+$ and $\sigma^-$) aligned along the magnetic field axis. The two beams are detuned relative to each other by twice the Zeeman splitting to compensate for the energy shift of hyperfine levels due to the quantization field\cite{Walhout_92}. We use this architecture for laser cooling, imaging, parity-projection, and qubit state initialization as shown schematically in Extended Data Fig.\,\ref{fig:suppfig_stateprep}a, with cooling and imaging performing comparably to the zero-field configuration.

To begin qubit preparation on atoms extracted into the preparation zone from the reservoir, an explicit parity-projection pulse is required since our loading mechanism typically results in more than one atom loaded per tweezer. Before ramping up SLM tweezers in the preparation zone, we perform PGC with light $60\,\mathrm{MHz}$ red-detuned from $F=2\rightarrow F'=3$ to induce pair-wise atom losses in AOD tweezers via light-assisted collisions\cite{Pampel_2025}. After 10\,ms, we handover atoms to the SLM array and observe roughly 50\% occupation and less than 1\% of sites with more than one atom. This parity-projection step therefore sets an upper bound on imaging survival and is crucial for obtaining well-separated imaging histograms. Additionally, we observe atoms trapped in weak out-of-plane potentials created by the SLM (Talbot effect), which we remove with a brief resonant push-out pulse\cite{Manetsch2024} applied to the entire array during PGC.

Fast, lossless imaging is then used to identify atoms for rearrangement.  We image for 10\,ms using PGC parameters, which achieves a site-resolved discriminant fidelity of $99.99\%$ (Extended Data Fig.\,\ref{fig:suppfig_stateprep}b) with a survival probability of $99.5\%$. Higher imaging fidelity and survival can always be achieved by imaging longer at larger detuning or weaker powers. Our 1D counter-propagating beam configuration offers the added benefit of a background-free imaging signal, eliminating the need for Fourier filtering since the beams do not directly scatter into the imaging objective. During imaging in particular, we observe that atoms from our lattice reservoir occasionally leak into the nearby preparation zone, which are then trapped into tweezers by imaging/cooling light. We avoid this by moving our lattice reservoir roughly a centimeter away from the objective field-of-view after every tweezer extraction cycle, and move it back before the next cycle. If not accounted for, atom spilling from the lattice and improper parity-projection in tweezers can each decrease imaging survival by $\sim\!5\%$, with the exact number dependent on reservoir density and imaging duration.

During the 20\,ms to 40\,ms of rearrangement, we apply laser cooling via electromagnetically-induced transparency (EIT)\cite{Kurtsiefer2024}. We use the same laser and beam geometry, except in a strong ``coupling" and weak ``probe" configuration, operating on a two-photon resonance $70\,\mathrm{MHz}$ blue-detuned from the $F = 2 \rightarrow F' = 2$ transition\cite{Kurtsiefer2024} with intensity ratio 13:1. We probe atom temperature via drop-recapture and extract radial temperatures of $12\,\upmu\mathrm{K}$ (Extended Data Fig.\,\ref{fig:suppfig_stateprep}c). A complementary adiabatic trap ramp-down measurement, which is more sensitive to axial temperatures, yields comparable results despite limited momentum projection of the cooling light along the tweezers' axial direction. All cooling lights (PGC and EIT) contain a repumper frequency addressing the $F = 1 \rightarrow F' = 2$ transition, created by modulating an electro-optic modulator at $\sim\!6.8\,\mathrm{GHz}$.\\

\noindent\textbf{Atom Rearrangement} \\
After stochastically loading the preparation zone $120 \times 12$ tweezer array, we rearrange the atoms into a defect-free array of typically 540 sites, except for the qubit flux demonstration in Fig.\,\ref{fig:fig1}c where we arrange to 600 sites. The same 2D AOD pair used for extracting atoms from the lattice reservoir performs rearrangement, controlled by a dedicated arbitrary waveform generator (AWG, Spectrum Instrumentation M4i.6631-x8). Leveraging our unique large-aspect-ratio preparation zone geometry, we execute efficient row-by-row sorting (Extended Data Fig.\,\ref{fig:suppfig_rearrange}a). In each row, a single parallel move fills all empty target sites using available atoms while navigating through inter-row gaps to avoid backbone SLM traps. Each row takes $700\,\upmu\mathrm{s}$ to sort, with EIT cooling active throughout.

To optimize real-time processing, we precompute all possible move segments as waveform chirps. During each experiment, the rearrangement program selectively synthesizes the precomputed waveform chirps based on the specific atom loading information of that shot (provided by a real-time image analysis program). We exploit the AWG's FIFO mode to reduce latency from waveform calculation. Since each row’s sorting is independent, we stream its waveform as soon as it is computed while moves for the next row are calculated in parallel. This reduces calculation latency by over an order of magnitude. Under optimal conditions, we allocate $\approx\!20\mathrm{ms}$ for rearrangement which takes into account data transfer latency, row-by-row sorting time, and a small buffer to accommodate program runtime fluctuations. For all sequences involving storage zone operation, we increase the camera ROI to include the storage zone, which increases latency (and therefore the time allocated for rearrangement) by $\sim\!10\text{-}20\,\mathrm{ms}$. This large ROI is unnecessary for future continuously operating experiments (e.g., for error-correcting circuits), where imaging is confined to the preparation zone and not required in the storage zone.

For 540-site rearrangement, we achieve an averaged target array filling fraction of 99.6\% (Extended Data Fig.\,\ref{fig:suppfig_rearrange}b), primarily limited by imaging survival. Given our vacuum lifetime of over 150\,s, losses from background gas collisions contribute less than 0.1\% to the rearrangement infidelity. Extended Data Fig.\,\ref{fig:suppfig_rearrange}a (lower) showcases a single-shot preparation zone fluorescence image of a defect-free array after rearranging to 600 sites. Atoms remaining outside the target array are automatically discarded when the subarray is transported to (re)load the storage zone. We intentionally rearrange into every other column of the SLM backbone array, creating a sparse array geometry that reduces the average atom travel distance during sorting, thereby improving both rearrangement fidelity and speed.\\

\noindent\textbf{Storage Array Building} \\
One of the key considerations in generating large-scale atom arrays is the trade-off between increasing tweezer spacing in order to minimize inter-tweezer crosstalk, and decreasing tweezer spacing for higher SLM diffraction efficiency. Therefore, our 3{,}240-site storage zone tweezer array features alternating horizontal spacings: wide, $6\,\upmu\mathrm{m}$-channels for minimal AOD-SLM tweezer crosstalk during atom transport through the array, and smaller spacings of $3\,\upmu\mathrm{m}$ in order to pack the array close to the SLM 0-th order where diffraction efficiency is significantly higher. The entire array is placed on one side (two quadrants) of the SLM 0-th order, as we empirically find that having large tweezer arrays in all four quadrants introduces additional ghost optical spots between tweezers.

The $90\times 36$-site storage array is divided into 6 interleaved subarrays of $45\times 12$ sites. Subarrays feature regular tweezer spacings of $9\,\upmu\mathrm{m}$ and are filled sequentially, such that the entire storage array is assembled in six iterations (see Extended Data Fig.\,\ref{fig:suppfig_rearrange}d and Fig.\,\ref{fig:fig2}c, inset). Within each iteration, we first load the preparation zone array stochastically, then rearrange atoms into a 540-site target array with a spacing of $9\,\upmu\mathrm{m}$ horizontally and $4.5\,\upmu\mathrm{m}$ vertically. Afterwards, atoms are picked up by AOD tweezers and transported to one of the six subarrays. During this transport, the subarray is vertically expanded from $4.5\,\upmu\mathrm{m}$ to match the $9\,\upmu\mathrm{m}$ spacing in the storage zone, while the horizontal spacing remains at $9\,\upmu\mathrm{m}$ (see also Supplementary Video 1). Maintaining identical horizontal spacings throughout the transport minimizes expansion overhead and enables faster transport. In addition, the sparse subarray structure largely avoids AOD heating at closer spacings.

We observe a slightly lower atom survival probability during tweezer transport to the storage zone when the static lattice reservoir is present, which can be recovered when the lattice potential itself is in motion simultaneously. As described previously, the lattice reservoir is translated by 1\,cm out of the objective field-of-view to avoid lattice spilling during high-contrast imaging, and moved back before the next tweezer extraction cycle. Therefore, we now simply time atom movement to the storage zone to occur synchronously with moving the lattice reservoir back into the objective field-of-view, thereby mitigating the above effect. Extended Data Fig.\,\ref{fig:suppfig_rearrange}c showcases storage zone assembly statistics after 300 repeated trials; on average, we load 3{,}193 atoms out of 3{,}240 sites, corresponding to a loading fraction of 98.5\%.\\

\noindent\textbf{Qubit Initialization} \\
After rearranging a defect-free array, we initialize the atoms into their qubit subspace. The qubit subspace is spanned by the two hyperfine clock states in the ground-state manifold of \(^{87}\)Rb, which we define as \(\ket{F=1, m_F=0} \equiv \ket{0}\) and \(\ket{F=2, m_F=0} \equiv \ket{1}\). To prepare the atoms in state \(\ket{0}\), we leverage the previously discussed 1D local beam configuration. Both of the counter-propagating preparation zone beams simultaneously address the \(F=1 \rightarrow F'=0\) and \(F=2 \rightarrow F'=2\) transitions. By selection rules, state $\ket{0}$ is dark to the $\sigma^\pm$ circularly polarized light field addressing \(F=1 \rightarrow F'=0\), while the states $\ket{F =1,m_{F}= \pm 1}$ are optically pumped to \(\ket{0}\) through $\ket{F' =0,m_{F'}= 0}$. Simultaneously, atoms in states \(F=2\) are depumped into \(F=1\) by addressing the \(F=2 \rightarrow F'=2\) transition (Extended Data Fig.\,\ref{fig:suppfig_stateprep}d). We observe a $1/e$-time of $5\,\upmu\mathrm{s}$ for populating state \(\ket{0}\). This technique for fast state initialization is advantageous as it requires only few scattered photons, minimizing heating from scattering, while simultaneously avoiding magnetic field rotations\cite{Evered2023}. From the preparation zone Rabi contrast, we infer a state preparation and measurement (SPAM) fidelity of \(98.1(3)\%\), likely limited by polarization impurities and off-resonant scattering to other hyperfine levels in the excited state. The state preparation fidelity can potentially be improved by incorporating Raman-assisted optical pumping schemes\cite{Evered2023}.\\

\noindent\textbf{Qubit Manipulation and Readout} \\
We drive the qubit states via optical Raman transitions, in a setup similar to previous works\cite{Levine2022} but operating 400\,GHz blue-detuned from the 780\,nm transition. The Raman beam drives the qubits at Rabi frequency $\mathrm\Omega / 2 \pi \approx 1\,\mathrm{MHz}$. At this intensity, we measure a $T_1$-like scattering lifetime of 10\,ms, in agreement with \textit{ab initio} Raman scattering calculations. From this, we infer a scattering-limited fidelity of 0.99995 per $\pi$-pulse. We note, however, that this represents an upper bound on our single-qubit gate fidelity; in practice, additional error sources such as atomic decoherence, intensity fluctuations, and phase noise may also contribute\cite{Manetsch2024}. The Raman beam is shaped to homogeneously address qubits in the large storage zone, which is achieved by using a fixed holographic phase plate (HOLO/OR ST-268) that forms a top-hat beam profile across the extent of the array. From measuring row-by-row Rabi frequency, we infer a beam homogeneity of approximately 1.04\% root-mean-square variation and 3.4\% peak-to-peak variation on the atoms (Extended Data Fig.\,\ref{fig:suppfig_raman1530}a). To minimize crosstalk between the Raman beam and the atoms in the preparation zone, we knife-edge the beam tail at the intermediate imaging plane of the beam shaping telescope, and thus remove residual light in the preparation zone.

To selectively read out qubits in $\ket{0}$, we apply a pushout pulse that resonantly removes atoms in $F = 2$ from the trap, then image atoms that remain in the $F=1$ ground-state manifold. To readout qubits in $\ket{1}$, we first apply a $\pi$-pulse to rotate the population to $\ket{0}$, followed by the same pushout and imaging pulse. We follow a slightly different imaging procedure depending on where atomic qubits are read out. In the preparation zone, we use PGC for qubit readout under a \textit{finite} magnetic field, which, in this work, is primarily used to identify occupied tweezer sites for atom rearrangement, but can also serve as mid-circuit readout in future error correction protocols. For global readout of all storage array qubits at the end of the experiment, we use a separate retro-reflected circularly-polarized global imaging beam at \textit{zero} magnetic field.\\

\noindent\textbf{Qubit Shielding} \\
Throughout the experiment and particularly during qubit preparation, we protect the $5S_{1/2}$ ground-state qubits in the storage zone from near-resonant photon scattering with the $5S_{1/2} \rightarrow 5P_{3/2}$ transition by addressing the $5P_{3/2} \rightarrow 4D_{5/2}$ transition near 1529\,nm\cite{Hu2024}. This results in an Autler-Townes splitting $\pm \upDelta_\mathrm{AT}\approx \pm 2\pi\times 10\,\mathrm{GHz}$ of the excited state, and therefore a suppression of scattering from nearby imaging/cooling light with detuning $\upDelta_\mathrm{cool}$ by a factor of $\sim\!(\upDelta_\mathrm{AT} / \upDelta_\mathrm{cool})^2 >\!10{,}000$ (``shielding").

The shielding light is sourced from a single-frequency fiber laser (Connet CoSF-D) outputting up to 10\,Watts at 1529.6\,nm, which is passed through an AOM (G\&H 3165-1) for fast switching control and fiber-coupled onto the experimental table. Similar to the Raman beam, we shape the shielding beam with a holographic phase plate (HOLO/OR ST-356) to create a flat-top beam profile of $\approx\!250\,\upmu\mathrm{m}$ along the vertical extent of the storage zone array with Gaussian beam waist $\approx\!50\,\upmu\mathrm{m}$ horizontally and approximately $2.5\,\mathrm{W}$ projected onto the atoms. The beam tails are knife-edged in an intermediate imaging plane to ensure no shielding crosstalk onto the preparation zone. The knife-edged flat-top beam profile is shown in Extended Data Fig.\,\ref{fig:suppfig_raman1530}b. 

In Fig.\,3c, we demonstrate the shielding effect in a spectroscopy scan by stepping the 1529\,nm-wavelength during simultaneous imaging and low-power shielding of storage zone atoms, and resolve the $4D_{5/2}$ resonance by suppression of global imaging signal. To further optimize shielding performance, we maximize the $T_2$-time of storage qubits as a function of shielding wavelength while locally imaging qubits in the preparation zone. In practice, the shielding light is operated free-running $\approx\!1\,\mathrm{GHz}$ red-detuned from the $5P_{3/2} \rightarrow 4D_{5/2}$ transition with a measured frequency stability of $\pm 10\,\mathrm{MHz}$.\\

\noindent\textbf{Maintaining Coherence while Reloading} \\
In Fig.\,\ref{fig:fig4}a, we apply N-repetitions of a XY16 dynamical decoupling (DD) pulse sequence (denoted as XY16-N) with fixed $\pi$-pulse spacing $2\tau \approx 1.6\,\mathrm{ms}$. During DD, we measure storage qubit coherence under various conditions: First, we quantify the effect of the distant MOT on storage qubits when pulsing the MOT at a 30\% duty cycle with the lattice reservoir lights off. This particular duty cycle is chosen to replicate typical MOT loading cycles in our qubit reloading protocol. Second, we additionally switch on the preparation zone imaging light while the reservoir is present for the entire probing duration to simulate the effect of concurrent local qubit preparation. We observe no difference between turning on local imaging light with atoms present in the preparation zone versus without, suggesting that the primary source of decoherence during qubit preparation arises from scattered light originating from the optics and apparatus, rather than from photons scattered by atoms in the preparation zone. Lastly, we shield storage zone qubits while the distant MOT is loaded and held at saturation, the lattice reservoir is present, and preparation zone imaging light is switched on for the entire probing duration. This simulates the most demanding application of storage array shielding. Complementary to Fig.\,\ref{fig:fig4}a which measures $T_2$-times under the conditions described above, we similarly probe depolarization of storage zone qubits when the qubit is initialized in either $\ket{0}$ (Extended Data Fig.\,\ref{fig:supp_t1t2}a) or $\ket{1}$ (Fig.\,\ref{fig:fig4}b). Here, in addition to the above variations, we also quantify storage qubit depolarization caused by the lattice reservoir light alone.\\

\noindent\textbf{Continuous Coherent Operation} \\
In this section, we detail the experimental sequence used to achieve our results of qubit reloading while maintaining storage qubit coherence shown in Figs.\,\ref{fig:fig3}b,c,d and Extended Data Figs.\,\ref{fig:quantumbattery}a,b. A sequence schematic is given in Extended Data Fig.\,\ref{fig:contcohseq}. First, we transport a lattice reservoir into the science region, from which tweezers repeatedly extract atoms into the preparation zone. Subsequently loaded lattice reservoirs are transported to the tweezer science region in parallel, such that the reservoir itself is replaced every two tweezer extractions. Once in the preparation zone, atoms undergo the qubit preparation sequence shown in Extended Data Fig.\,\ref{fig:suppfig_stateprep}a before being transported into the storage zone. The complete qubit preparation sequence, including the move to the storage array, takes a total of $\sim\!80\,\mathrm{ms}$ and constitutes one reloading cycle. After initial assembly of the storage qubit array, we continue preparing newly state-initialized qubit ensembles in the preparation zone, and eject and refill one of six qubit subarrays in the storage zone as described in the main text. To eject a subarray, the qubits are transferred back into overlapped AOD tweezers and accelerated out of the objective field-of-view. Storage subarray ejection takes $\approx\!5\,\mathrm{ms}$ and occurs in parallel to the preparation zone image used for atom rearrangement. For Fig.\,\ref{fig:fig3}b, we loop this qubit reloading sequence for variable time before our global imaging readout. Shielding light is applied to the storage zone during the entire experimental sequence.

For Figs.\,\ref{fig:fig3}c,d, we additionally apply dynamical decoupling sequences $(\text{X}_{\pi/2}-\text{XY16-64}-\text{X}_{- \pi/2})$ with a fixed $\pi$-pulse spacing $2\tau \approx 1.1\,\mathrm{ms}$ onto all storage zone qubits during each reloading cycle. Therefore, within each loop, qubits are first rotated $\ket{0} \rightarrow \ket{+}$, then undergo the XY16 decoupling sequence to mitigate dephasing. Right before fresh qubits are moved into the storage array from the preparation zone, we apply a $\text{X}_{- \pi/2}$ pulse to map remaining coherence to population by rotating back to state $\ket{0}$, replace the oldest qubit subarray, then rotate all qubits again to $\ket{+}$ as a new reloading cycle starts. This is repeated for a variable number of times, where each subarray is exchanged with a set of fresh qubits every six iterations. In Extended Data Fig.\,\ref{fig:quantumbattery}a, we supplement main text Fig.\,\ref{fig:fig3}c by averaging readout probability over the entire storage array instead of analyzing each subarray individually.

Instead of mapping coherence back to $\ket{0}$ population after each reloading cycle, we can also map it to alternating $\ket{0}$ or $\ket{1}$ population by choosing $(\text{X}_{\pi/2}-\text{XY16-64}-\text{X}_{+\pi/2})$ as the cyclic decoupling sequence. This results in subarrays 1, 3, and 5 hosting qubits in superposition state $\ket{+}$ and subarrays 2, 4, and 6 qubits in the opposite state $\ket{-}$ during dynamical decoupling after initial array assembly. When mapped back to population, this yields a checkerboard pattern of qubit states $\ket{1}$ and $\ket{0}$ in the storage array. Analogous to Fig.\,\ref{fig:fig3}d, this is shown in Extended Data Fig.\,\ref{fig:quantumbattery}b where the storage array is read out in different qubit bases and each subarray analyzed individually.\\

\noindent\textbf{Control System and Timing} \\
We use National Instruments (NI) cards to generate digital and analog control signals (NI PXIe-6535 and NI PXIe-6738, respectively) for the laser cooling and trapping stages of our experiment, including the MOT loading, dual-lattice transport, and qubit preparation sequences. For operations that require waveform generation with nanosecond precision, we utilize arbitrary waveform generators (AWGs, Spectrum Instrumentation DN2.663-04 and M4i.6631-x8) whose output is triggered by the NI cards. In our experiments, AWGs handle timing of single-qubit gates and dynamical decoupling pulses, and supply the chirped waveforms for atom sorting and transport in AOD tweezers.

Our control system is designed to allow for practically unlimited duration of continuous operation. For NI-generated control signals, we calculate and stream the waveform samples on-the-fly to circumvent memory limitations. For AWG-controlled atom transport and dynamical decoupling, we instead store the precalculated waveform in the on-board memory and loop over it for an arbitrary number of times. Here, the dynamical decoupling sequence requires particular care in selecting the signal frequency when looping over the same memory segment as to avoid phase jumps in the IQ-modulated 6.8\,GHz microwave signal. More details of our control software will be discussed in an independent manuscript which is currently in preparation.

Calibrating and maintaining intensity of light pulses that are too short for active real-time stabilization is a significant technical challenge in continuously operating experiments. For our optical Raman light, we employ an FPGA-based digital servo with digital sample-and-hold\cite{Neuhaus2024} to stabilize pulse intensity, which eliminates analog hold decay and integral windup, and enables calibration pulses as short as $5\,\upmu\mathrm{s}$. This calibration pulse is inserted before every XY16-64 decoupling cycle with the 6.8\,GHz microwave source detuned by $20\,\mathrm{MHz}$ to ensure that qubit states are not driven. Although the calibration pulse flashes onto existing storage qubits, its duration is three orders of magnitude shorter than the $\sim\!10\,\mathrm{ms}$ $T_1$-like scattering timescale associated with the Raman light. With the digital hold, we actively stabilize every decoupling cycle \textit{in-situ}, and achieve no long-term decay in pulse intensity and pulse-to-pulse error of $\lesssim\!1\%$.\\

\noindent\textbf{Data Analysis} \\
For the measurements in Figs.\,\ref{fig:fig4}a,b, Fig.\,\ref{fig:fig3}b, and Extended Data Fig.\,\ref{fig:supp_t1t2}a, we read out the qubit state after the probing duration and define contrast as $(P_{{0}}(t) - P_{{1}}(t)) / (P_{a}(t) - P_{\ket{m_F =\pm 1}}(t=0))$. Here, $P_{{0}}(t)$ ($P_{{1}}(t)$) denotes the probability to measure qubits in $\ket{0}$ ($\ket{1}$) at given time $t$ by reading out the $F=1$ hyperfine level without (with) a preceding $\pi$-pulse as described above. $P_{a}(t)$ is the probability to measure an atom in any state by omitting the pushout pulse before imaging atoms (lifetime measurement). Additionally, we correct for qubits initially populating neighboring Zeeman states $\ket{F=1, m_F = \pm 1}$ at $t=0$ denoted as $P_{\ket{m_F = \pm 1}}(t=0)$, since we observe $5\text{-}10\%$ leakage from state $\ket{0}$ into other $m_F$ states within the $F = 1$ manifold when transporting qubits from the preparation to the storage array using AOD tweezers. This is attributed to beating of radio frequencies driving the transitions between $\ket{0} \rightarrow \ket{F=1, m_F = \pm1}$ levels, which can be mitigated by operating at higher quantization fields or fine-tuning the radio frequencies applied to the AODs in future experiments. To measure $P_{\ket{m_F = \pm 1}}(t=0)$, we isolate qubits in states $\ket{F=1, m_F = \pm 1}$ by first applying a pushout pulse, then a $\pi$-pulse to transfer the population $\ket{0}\rightarrow\ket{1}$, followed by a final pushout pulse before readout. For all measurements of Figs.\,\ref{fig:fig4}a,b and Extended Data Fig.\,\ref{fig:supp_t1t2}a, we fit an exponential decay to the contrast and quote the fitted $1/e$ decay time as $T_2$- and $T_1$-times, which are presented in Extended Data Fig.\,\ref{fig:supp_t1t2}b.

For the measurements in Figs.\,\ref{fig:fig3}c,d and Extended Data Fig.\,\ref{fig:quantumbattery}a,b, ``Readout Probability" is defined as $(P_{{0}}(t) - P_{\ket{m_F =\pm 1}}(t=0)) / (P_{a}(t) - P_{\ket{m_F=\pm 1}}(t=0))$, where $P_{{0}}(t)$ again denotes the probability to measure qubits in $\ket{0}$ at given time $t$ as described above, and $P_{a}(t)$ the lifetime measurement. For the red-shaded lines in Fig.\,\ref{fig:fig3}c and Extended Data Fig.\,\ref{fig:quantumbattery}a, we apply a $\pi$-pulse before the measurement. As before, we measure and correct for qubits initially populating neighboring Zeeman states $\ket{F=1, m_F = \pm 1}$ due to state leakage during atom transport.\\

\noindent\textbf{Data Availability} \\
\noindent The data that support the findings of this study are available from the corresponding authors on request.


\end{document}